\documentclass[usenatbib]{mnras}
\usepackage{newtxtext,newtxmath}
\usepackage[T1]{fontenc}
\usepackage{ae,aecompl}
\usepackage{float}
\usepackage{comment}
\usepackage{color}
\usepackage{caption}
\usepackage{multirow}
\usepackage{docmute}
\usepackage{booktabs}
\usepackage{graphicx}	
\usepackage{amsmath}	
\usepackage{amssymb}
\usepackage{mathtools}
\usepackage[dvipsnames]{xcolor}
\usepackage{siunitx}
\usepackage{pifont}
\usepackage{tikz}
\usepackage{threeparttable}

\newcommand{\cmark}{\ding{51}}%
\newcommand{\xmark}{\ding{55}}%

\hypersetup{
  linkcolor  = blue,
  citecolor  = blue,
  urlcolor   = magenta,
  colorlinks = true,
}

\definecolor{lime}{HTML}{A6CE39}
\DeclareRobustCommand{\orcidicon}{%
	\begin{tikzpicture}
	\draw[lime, fill=lime] (0,0) 
	circle [radius=0.16] 
	node[white] {{\fontfamily{qag}\selectfont \tiny ID}};
	\draw[white, fill=white] (-0.0625,0.095) 
	circle [radius=0.007];
	\end{tikzpicture}
	\hspace{-2mm}
}
\foreach \x in {A, ..., Z}{%
	\expandafter\xdef\csname orcid\x\endcsname{\noexpand\href{https://orcid.org/\csname orcidauthor\x\endcsname}{\noexpand\orcidicon}}
}

\title[Mass Estimates from Proper Motions]{Accurate mass estimates from the proper motions of dispersion-supported galaxies} 

\author[Lazar \& Bullock]{Alexandres Lazar\thanks{\href{mailto:aalazar@uci.edu}{aalazar@uci.edu}} and
James S. Bullock\orcidB{}\thanks{\href{mailto:bullock@uci.edu}{bullock@uci.edu}}
\\
Department of Physics and Astronomy, University of California, Irvine, CA 92697 USA
}

\date{Accepted 2020 March 6. Received 2020 March 6; in original form 2019 July 19}
\pubyear{2020}

\begin{document}
\label{firstpage}
\pagerange{\pageref{firstpage}--\pageref{lastpage}}
\maketitle

\begin{abstract}
We derive a new mass estimator that relies on internal proper motion measurements of dispersion-supported stellar systems, one that is distinct and complementary to existing estimators for line-of-sight velocities. Starting with the spherical Jeans equation, 
we show that there exists a radius where the mass enclosed depends only on the projected tangential velocity dispersion, assuming that the anisotropy profile slowly varies. This is well-approximated at the radius where the log-slope of the stellar tracer profile is $-2$: $r_{-2}$. The associated mass is $M(r_{-2}) = 2 G^{-1} \langle \sigma_{\mathcal{T}}^{2}\rangle^{*} r_{-2}$ and the circular velocity is $V^{2}({r_{-2}}) = 2\langle \sigma_{\mathcal{T}}^{2}\rangle^{*}$. For a Plummer profile $r_{-2} \simeq 4R_e/5$. Importantly, $r_{-2}$ is smaller than the characteristic radius for line-of-sight velocities derived by Wolf et al. 2010. Together, the two estimators can constrain the mass profiles of dispersion-supported galaxies. We illustrate its applicability using published proper motion measurements of dwarf galaxies Draco and Sculptor, and find that they are consistent with inhabiting cuspy NFW subhalos of the kind predicted in CDM but we cannot rule out a core. We test our combined mass estimators against previously-published, non-spherical cosmological dwarf galaxy simulations done in both CDM (naturally cuspy profile) and SIDM (cored profile). For CDM, the estimates for the dynamic rotation curves are found to be accurate to $10\%$ while SIDM are accurate to $15\%$. Unfortunately, this level of accuracy is not good enough to measure slopes at the level required to distinguish between cusps and cores of the type predicted in viable SIDM models without stronger priors. However, we find that this provides good enough accuracy to distinguish between the normalization differences predicted at small radii ($r \simeq r_{-2} < r_{\rm core}$) for interesting SIDM models. As the number of galaxies with internal proper motions increases, mass estimators of this kind will enable valuable constraints on SIDM and CDM models.
\end{abstract}


\section{Introduction}
The $\Lambda$CDM cosmogony, while successful in describing the large scale structure of our universe, still suffers from potential discrepancies in modeling the properties on small scales, primarily for dark matter halos that are expected to host the observed dwarf galaxies.
For example, Milky-Way satellites have significantly lower dark matter densities in the inner regions compared to the corresponding subhalos in cosmological $N$-body simulations --- this is known as the {\it Too Big To Fail} problem \citep{boylan2011too}. A potentially related issue concerns the inner dark matter density profiles inferred from the rotation curves of small disk galaxies, many of which are observed to be cored/flat, while simulated $\Lambda$CDM halos are cusped/rising --- this is the {\it Cusp-core} problem \citep{flores1994,moore1994,de2010core}. Feedback from star formation can potentially explain this discrepancy in larger dwarf galaxies \citep{governato2010bulgeless,pontzen2012supernova}. However, if dark matter cores exist within galaxies that have had too little star formation ($M_\star \lesssim 10^{6}\ M_\odot$) to affect the dark matter density slopes \citep{di2013dependence,chan2015impact,tollet2016nihao}, then this could be an indication that the dark matter is something other than CDM \citep[see][and references there in]{bullock2017small}.

Though particularly important, the question of whether or not the smallest galaxies have cusps or cores is notoriously difficult to answer owing to the fact that they are dispersion supported. While it is possible to quantify the detailed mass profiles of spheroidal galaxies through the use of kinematic measurements of individual stars in 3D \citep[e.g.][]{wilkinson2002dark,strigari2007astrometry}, until recently we have been limited to data sets that include only 1D velocities along the line-of-sight.  This introduces a degeneracy between the inferred mass profile slope and the underlying velocity dispersion anisotropy parameter $\beta$, which quantifies the intrinsic difference between the radial and tangential velocity dispersions. 

One robust measurement that is possible with line-of-sight velocities is the integrated mass within a single characteristic radius for each galaxy. This idea was first emphasized by \cite{walker2009universal}, who used spherical Jeans modelling to show that the integrated mass within an effective radius was independent of assumed $\beta$ for a wide variety of assumptions for many galaxies. \cite{wolf2010accurate} extended this idea, also using Jeans modeling, to show that there exists, analytically, an idealized radius within which the mass inferred from line-of-sight velocities is formally insensitive to $\beta$. Under mild assumptions, this radius is where the log-slope of the stellar tracer profile is equal to $-3$. Both the Walker and Wolf mass estimators do remarkably well when compared to \textit{ab initio} cosmological simulations of (non-spherical) dwarf galaxies in \citep{campbell2017knowing,gonzalez2017dwarf}. They are also used extensively to interpret observed line-of-sight velocity dispersion measurements \citep[see][and references there in]{simon2019}.

We are entering a new era of astrometry, such that the internal proper motions in distant dwarf spheroidal galaxies are now becoming possible to measure with the advent of {\tt GAIA} \citep{brown2016gaia,prusti2016gaia,brown2018gaia,helmi2018gaia}. Additionally, {\tt LSST} may provide similar advances \citep{lsst2009}. Measurements of stellar velocities along the plane-of-the-sky promise an important new window into the mass and density structure of dwarf galaxies.  The results of \cite{massari20173d} and \cite{massari2019stellar} provide an exciting first look at what we expect to measure in the coming years by providing  plane of the sky velocity dispersion measurements for Sculptor and Draco, respectively.

The article is outlined as follows: In Section~\ref{sec:preliminaries}, we briefly introduce the spherical Jeans equation and the coordinate system used as the basis of our analysis. Section~\ref{sec:estimators} derives the mass estimators by combining the Jeans equation and proper motions measured from the plane of the sky, which includes the key assumptions considered therein. Section~\ref{sec:observation.results} demonstrates the use of the combined mass estimators to provide an implied mass-density slope for currently available proper motions of Draco and Sculptor. Section~\ref{sec:mock.observations} assesses our estimators with mock observations constructed from high-resolution simulations. In Section~\ref{sec:discussion}, we discuss possible biases that might arise due to Jeans modelling, and finally, Section~\ref{sec:conclusion} summarizes our results and we provide concluding remarks.

\section{Prelimnaries}
\label{sec:preliminaries}
In what follows, lower case $r$ represents the (physical) three-dimensional radius and the upper case $R$ represents the (physical) two-dimensional projected radius.

\subsection{The Spherical Jeans Equation}
\label{sec:spherical.jean.equation}
For a spherically symmetric steady-state system, the first moment of the collisionless Boltzmann equation for a stellar phase-space distribution takes the form of the spherical Jeans equation \citep{binney2011galactic}:
\begin{align}
    -\frac{d\Phi(r)}{dr} 
    = 
    \frac{1}{n_{\star}(r)}\frac{d}{dr}\left(n_{\star} \sigma_{r}^{2}(r)\right) 
    + \frac{2\beta\sigma^{2}_{r}(r)}{r} \, ,
    \label{eq:spherical_jeans}
\end{align}
which relates the total gravitational potential, $ \Phi(r)$, of the stellar system to its two tracers: the intrinsic radial velocity dispersion, $\sigma_{r}^{2} := \langle v_{r}^{2}\rangle - \langle v_{r} \rangle^{2}$, and the three-dimensional stellar number density, $n_{\star}(r)$. The quantity,
\begin{align}
    \beta(r) 
&   := 
    1 - \frac{\sigma_{\theta}^{2} + \sigma_{\phi}^{2}}{2\sigma_{r}^{2}} \, ,
\end{align}
is a measure of the velocity dispersion {\it anisotropy} of the tracer population, where $\sigma_{\theta}$ and $\sigma_{\phi}$ are the intrinsic velocity dispersion tangential to radius $r$. We will assume that $\sigma_{\theta} = \sigma_{\phi}$.
Radially biased systems tend to have $\beta \rightarrow 1$ while $\beta \rightarrow -\infty$ constitutes tangentially biased measurements. In addition, the total intrinsic velocity dispersion follows
\begin{align}
    \sigma_{\rm tot}^{2}(r)
    = \sigma_{r}^{2} + \sigma_{\theta}^{2} + \sigma_{\phi}^{2} 
    = (3-2\beta)\sigma_{r}^{2}(r)
    \, .
    \label{eq:total.dispersion}
\end{align} 

The total mass profile of the dynamical system is an implied quantity of Eq.~\eqref{eq:spherical_jeans}, such that,
\begin{align}
    M(r|\beta)
    =
    \frac{r \sigma_{r}^{2}(r)}{G}
    \left( \gamma_{\star} + \gamma_{\sigma} - 2\beta \right) 
    \, ,
    \label{eq:jeans_mass}
\end{align}
where $G$ is Newtons gravitational constant and the logarithmic slopes are defined as
\begin{align}
    \gamma_{\star} := -\frac{d\log n_{\star}}{d\log r}
    ;&\qquad 
    \gamma_{\sigma} := -\frac{d\log \sigma_{r}^{2}}{d\log r};
    \\ \nonumber 
    \gamma_{\beta} &:= -\frac{d\log \beta}{d\log r}
    \, .
\end{align}

\subsection{Coordinate System of Measurements}
\label{sec:coordinates}
We use the coordinate system discussed in \cite{strigari2007astrometry} such that the three-dimensional components of stars' velocity in a spherically, steady-state systems are comprised of the components radial, $v_{r}$, and transverse, $v_{\theta}$ and $v_{\phi}$. The projected proper motions are composed of these three-dimensional velocities, that is, along the measured {\it line-of-sight},
\begin{align}
    v_{\rm los} 
    &= v_{r}\cos\theta + v_{\theta}\sin\theta
    \, ,
\end{align}
where $\vec{r}\cdot \vec{z} = \cos\theta$ and $\vec{z}$ is the line-of-sight direction, and along the plane of the sky, the components {\it parallel} and {\it transverse} to the projected radius $R$ are
\begin{align}
    v_{\mathcal{R}} 
    &= v_{r}\sin\theta + v_{\theta}\cos\theta
    \,\quad \mathrm{and}\quad 
    v_{\mathcal{T}} 
    = v_{\phi}
    \, ,
\end{align}
respectively. 
Here, the variances of the velocity dispersions are given by $\sigma_{i}^{2} := \langle v_{i}^{2} \rangle$ and $\sigma_{\phi} = \sigma_{\theta}$ is assumed. 
The derived mapping of the observable proper motions to the deprojected, three-dimensional tracer profiles are
\begin{align}
    \Sigma_{\star} \sigma_{\rm los}^{2}(R)
    &=
    \int_{R^{2}}^{\infty}
    \frac{dr^{2}}{\sqrt{r^{2} - R^{2}}}
    \left[
    1- \frac{R^{2}}{r^{2}}\beta
    \right]
    n_{\star}\sigma_{r}^{2}\, ,
    \label{eq:sig_los.mapping}
    \\
    \Sigma_{\star} \sigma_{\mathcal{R}}^{2}(R)
    &=
    \int_{R^{2}}^{\infty}
    \frac{dr^{2}}{\sqrt{r^{2} - R^{2}}}
    \left[
    1 - \beta + \frac{R^{2}}{r^{2}}\beta
    \right]
    n_{\star}\sigma_{r}^{2} \, ,
    \label{eq:sig_projR.mapping}
    \\
    \Sigma_{\star} \sigma_{\mathcal{T}}^{2}(R)
    &=
    \int_{R^{2}}^{\infty}
    \frac{dr^{2}}{\sqrt{r^{2} - R^{2}}}
    \left[ 1-\beta\right]
    n_{\star}\sigma_{r}^{2} \, .
    \label{eq:sig_projT.mapping}
\end{align}
The combination of the proper motions also satisfy $\sigma_{\rm tot}^{2} = \sigma_{\mathcal{R}}^{2} + \sigma_{\mathcal{T}}^{2} + \sigma_{\rm los}^{2}$. For an observed galaxy, $\Sigma_{\star}(R)$ is the projected stellar density, which is related to the three-dimensional $n_{\star}(r)$ via an Abel inversion, Eq~\eqref{eq:abel.inversion}. For the relevance of the proceeding text, we will focus on the additional constraint imposed by 

\section{Mass Estimators from Proper Motions}
\label{sec:estimators}
In this section, quantities enclosed in brackets with an asterisk as $\langle\cdots\rangle^{*}$ indicates a measurement to be {\it luminosity-weighted}, $r_{1/2}$ is the three-dimensional, deprojected half-light radius, and $R_{e}$ is the two-dimensional effective radius.


\renewcommand{\arraystretch}{1.2}%

\begin{table*}
    \centering
    \caption{ 
        Observational measurements considered in this analysis.
    }
    \begin{threeparttable}
    \begin{tabular}{cccccc|cc|cc} 
        \toprule
        Galaxy & 
        {$\sqrt{ \langle\sigma_{\rm los}^{2} \rangle^{*}}$} & {$\sqrt{\langle\sigma_{\mathcal{R}}^{2}\rangle^{*}}$} & {$\sqrt{\langle \sigma_{\mathcal{T}}^{2}\rangle^{*}}$} 
        & {$R_{e}$} & {$r_{\rm 1/2}$} 
        & {$M_{-3}$} & {$V_{-3}$} 
        & {$M_{-2}$} & {$V_{-2}$}
        \\
        & {$[\rm km\ s^{-1}]$} & 
        {$[\rm km\ s^{-1}]$} & 
        {$[\rm km\ s^{-1}]$} & 
        {$[\rm pc]$} & {$[\rm pc]$} 
        & {$[M_{\odot}]$} 
        & {$[\rm km\ s^{-1}]$} 
        & {[$M_{\odot}]$} 
        & {$[\rm km\ s^{-1}]$}
        \\ \midrule
        Draco & 
        {${}^{(a,b)} 10.1^{+ 0.5}_{-0.5}$} & 
        {${}^{(e)} 11.0^{+ 2.1}_{- 1.5}$} & 
        {${}^{(e)} 9.9^{+ 2.3}_{- 3.1}$} & 
        {${}^{(d,f)} 214^{+2}_{-2}$} & {$ 279^{+2}_{-2}$}
        & {$2.03^{+0.20}_{-0.20}\times 10^{7}$} 
        & {$17.5^{+0.9}_{-0.9}$}  
        & {$7.80^{+4.89}_{-3.63} \times 10^{6}$} 
        & {$14.0^{+3.3}_{-4.4}$}
        \\ 
        Sculptor & 
        {${}^{(a,b)} 9.0^{+ 0.2}_{-0.2}$} & 
        {${}^{(c)} 11.5^{+ 4.3}_{-4.3}$} & 
        {${}^{(c)} 8.5^{+ 3.2}_{-3.2}$} & 
        {${}^{(d,f)} 280^{+1}_{-1}$} & {$365^{+1}_{-1}$}
        & {$2.11^{+0.09}_{-0.09} \times10^{7}$} 
        & {$15.6^{+0.3}_{-0.3}$}  
        & {$7.53^{+5.67}_{-5.67} \times 10^{6}$}
        & {$12.0^{+4.5}_{-4.5}$}
        \\ \bottomrule
    \end{tabular}
    \end{threeparttable}
    \begin{tablenotes}
    \item[]
     Here, $M_{-3}$ and $V_{-3}$ are computed using Eqs. \eqref{eq:wolf_mass} and \eqref{eq:wolf_vcirc}, respectively, while $M_{-2}$ and $V_{-2}$ used Eqs. \eqref{eq:lazar_mass} and \eqref{eq:lazar_vcirc}, respectively.
     \item[]
    {\bf \emph{References}} -- 
        $(a)$: \protect\cite{walker2009universal}, 
        $(b)$: \protect\cite{wolf2010accurate}, 
        $(c)$: \protect\cite{massari20173d}, 
        $(d)$: \protect\cite{munoz2018megacam},
        $(e)$: \protect\cite{massari2019stellar}, 
        $(f)$: \protect\cite{simon2019}.
    \end{tablenotes}
    \label{tab:1}
\end{table*}


\subsection{Measurements along the Line-of-sight}
Here, we rederive the main results from \cite{wolf2010accurate} using the assumptions discuss there-in: Consider a velocity dispersion-supported stellar system that is well studied, such that $\Sigma_{\star}(R)$ and $\sigma_{\rm los}(R)$ are determined accurately by observations. In this system, all of the stars are assumed to be bound with no dynamical interlopers. 
If we model the systems mass profile using the Jeans equation, any viable solution will keep the combination of $\Sigma_{\star}\sigma_{\rm los}^{2}(R)$ fixed to within allowable errors. 
We start with the mapping of $\sigma_{\rm los}$ to $\sigma_{r}$. To utilize Eq.~\eqref{eq:spherical_jeans}, Eq.~\eqref{eq:sig_los.mapping} is massaged to an invertable form that is applicable to that of an Abel inversion:
\begin{align}
    \Sigma_{\star}\sigma_{\rm los}^{2}(R)
    &=
    \int_{R^{2}}^{\infty}
    \frac{dr^{2}}{\sqrt{r^{2} - R^{2}}}
    \left[
    n_{\star}\sigma_{r}^{2}(1-\beta)
    +
    \int_{r^{2}}^{\infty} d\tilde{r}^{2} \frac{\beta n_{\star} \sigma_{r}^{2}}{2\tilde{r}^{2}}
    \right] \, .
    \nonumber 
    \label{eq:sig_los.mapping.inv}
\end{align}
The term in the brackets on the right-hand side has to be a well-defined quantity, as the left-hand side is an accurate, observed quantity ignorant of the form of $\beta$. Therefore, we are allowed to compare different forms of $\beta$ with one another; we equate the isotropic form of the integrand, $\beta = 0$, with an integrand that is dependent on some arbitrary $\beta$, as this is the simplest case one can consider as a comparison: 
\begin{align}
    n_{\star}\sigma_{r}^{2}\big|_{\beta = 0} 
    &= 
    n_{\star}\sigma_{r}^{2}(1-\beta)  + \int_{r^{2}}^{\infty} d\tilde{r}^{2} \frac{\beta n_{\star} \sigma_{r}^{2}}{2\tilde{r}^{2}}
    \, .
\end{align}
By then taking a derivative in respect to $\log r$ and introducing a factor of $r\sigma_{r}^{2}/G$ on both sides, we can massage the left-hand and right-hand side into their respective forms of Eq.~\eqref{eq:jeans_mass} and evaluate the difference:
\begin{align}
    M(r|\beta) - M(r|0)
    &=
    \frac{r \sigma_{r}^{2} \beta}{G} 
    \left(\gamma_{\star} + \gamma_{\sigma} + \gamma_{\beta} - 3 \right)
    \, .
    \label{eq:los_mass_diff}
\end{align}
From this expression, we see that there can exist a radius, $r_{\rm eq}$, where the term in the parentheses vanishes based off of mapping projected line-of-sight measurements to the intrinsic quantities of the system, that is, 
\begin{align}
    \gamma_{\star}(r_{\rm eq}) 
    &= 
    3 - 
    \gamma_{\sigma}(r_{\rm eq}) - 
    \gamma_{\beta}(r_{\rm eq})
    \, .
    \label{eq:wolf_constraint}
\end{align}
Moreover, if $\sigma_{r}^{2}(r)$ and $\beta(r)$ are slowly varying, such that the log-slope profiles are approximately zero, i.e., $\gamma_{\sigma}(r_{\rm eq}) \simeq 0$ and $\gamma_{\beta}(r_{\rm eq}) \simeq 0$, then the degeneracy of the mass profile written in Eq.~\eqref{eq:jeans_mass} is effectively minimized. This would then have the right-hand side of Eq.~\eqref{eq:los_mass_diff} to be subsequently null. Furthermore, if $\gamma_{\star}(r_{\rm eq}) \simeq 3$, then this equates the radius of minimized anisotropy as $r_{\rm eq} \simeq r_{-3}$, which is the radius where the differential log-gradient of the stellar tracer profile is $-3$. 

To determine the value of $M(r_{\rm eq})$, Eq.~\eqref{eq:sig_los.mapping.inv} is deprojected via an Abel inversion to isolate out the combination of $n_{\star}\sigma_{r}^{2}$ \citep[Eq. A5;][]{wolf2010accurate}. This is then hit with a derivative in respect to $\log r$ and is inserted into Eq.~\eqref{eq:jeans_mass} to obtain
\begin{align}
    \frac{GM(r)}{r}
    &=
    (3-2\beta)\sigma_{r}^{2}(r)
    +
    \left( \gamma_{\star} + \gamma_{\sigma} - 3 \right)\sigma_{r}^{2}(r)
    \nonumber \\
    &=
    \sigma_{\rm tot}^{2}(r)
    +
    \left( \gamma_{\star} + \gamma_{\sigma} - 3 \right)\sigma_{r}^{2}(r)
    \, ,
\end{align}
where we have related the total intrinsic velocity dispersion using Eq~\eqref{eq:total.dispersion}. From the assumptions prior, if $\gamma_{\sigma}(r_{\rm eq}) \ll 3$, then the parenthetical term vanishes and $r_{\rm eq}\simeq r_{-3}$, giving us
\begin{align}
    M(r_{-3}) 
    \simeq
    \frac{\sigma_{\rm tot}^{2}(r_{-3}) r_{-3}}{G}
    \, .
\end{align}
 \cite{wolf2010accurate} showed that to a good approximation, $\sigma_{\rm tot}^{2}(r_{-3}) \simeq \langle \sigma_{\rm tot}^{2} \rangle^{*}$ for models that match observations. Furthermore, spherical symmetry demands that the line-of-sight dispersion obeys $\langle \sigma_{\rm tot}^{2} \rangle^{*} = 3\langle \sigma_{\rm los}^{2} \rangle^{*}$ (see Section \ref{sec:virial}). This will lead us to obtain an {\it idealized} estimator at $r_{-3}$:\footnote{Throughout, we refer to an {\it idealized} solution as one that considers the quintessential case of $\gamma_{\beta} = \gamma_{\sigma} = 0$ at the radius that minimizes the anisotropy. We do not expect physical results to perfectly match this behavior, but we instead presume the profiles to be relatively small enough at the expected radius where this is prominent. We will remain agnostic on this point until later in the article.}
\begin{align}
    M_{-3}^{\rm ideal} 
    &\equiv M(r_{-3}) 
    =
    \frac{3\langle \sigma_{\rm los}^{2} \rangle^{*} r_{-3}}{G} 
    \, .
    \label{eq:wolf_mass}
\end{align} 
Additionally, with the foundations of spherical symmetry, the implied circular velocity at $r_{-3}$ is particularly simple
\begin{align}
    V_{\rm circ}(r_{-3})
    &=
    \sqrt{\ 3 \langle \sigma_{\rm los}^{2} \rangle^{*}} 
    \, .
    \label{eq:wolf_vcirc}
\end{align} 
 \cite{wolf2010accurate} showed that for a variety of analytical stellar profiles, $r_{-3}$ is close to $r_{1/2} \simeq 4R_{e}/3$, giving
\begin{align}
    M(r_{-3}) &\simeq 
    \frac{3\langle \sigma_{\rm los}^{2} \rangle^{*} r_{-3}}{G} 
    ;\qquad
    M(4R_{e}/3) \simeq 
    \frac{4\langle \sigma_{\rm los}^{2} \rangle^{*} R_{e}}{G}
    \, .
\end{align} 
In the coming sections, we utilize the arguments stipulated here in the derivation of Eq.~\eqref{eq:wolf_mass}, where we seek to determine if another radius, one that is also independent of the anisotropy, exists for the two other proper motion mappings, such that it is independent of $r_{\rm eq}$ found previously. From here-on, we will refer Eq.~\eqref{eq:wolf_mass} as $M_{-3}$ and Eq.~\eqref{eq:wolf_vcirc} as $V_{-3}$.

\begin{figure}
    \centering
    \includegraphics[width=\columnwidth]{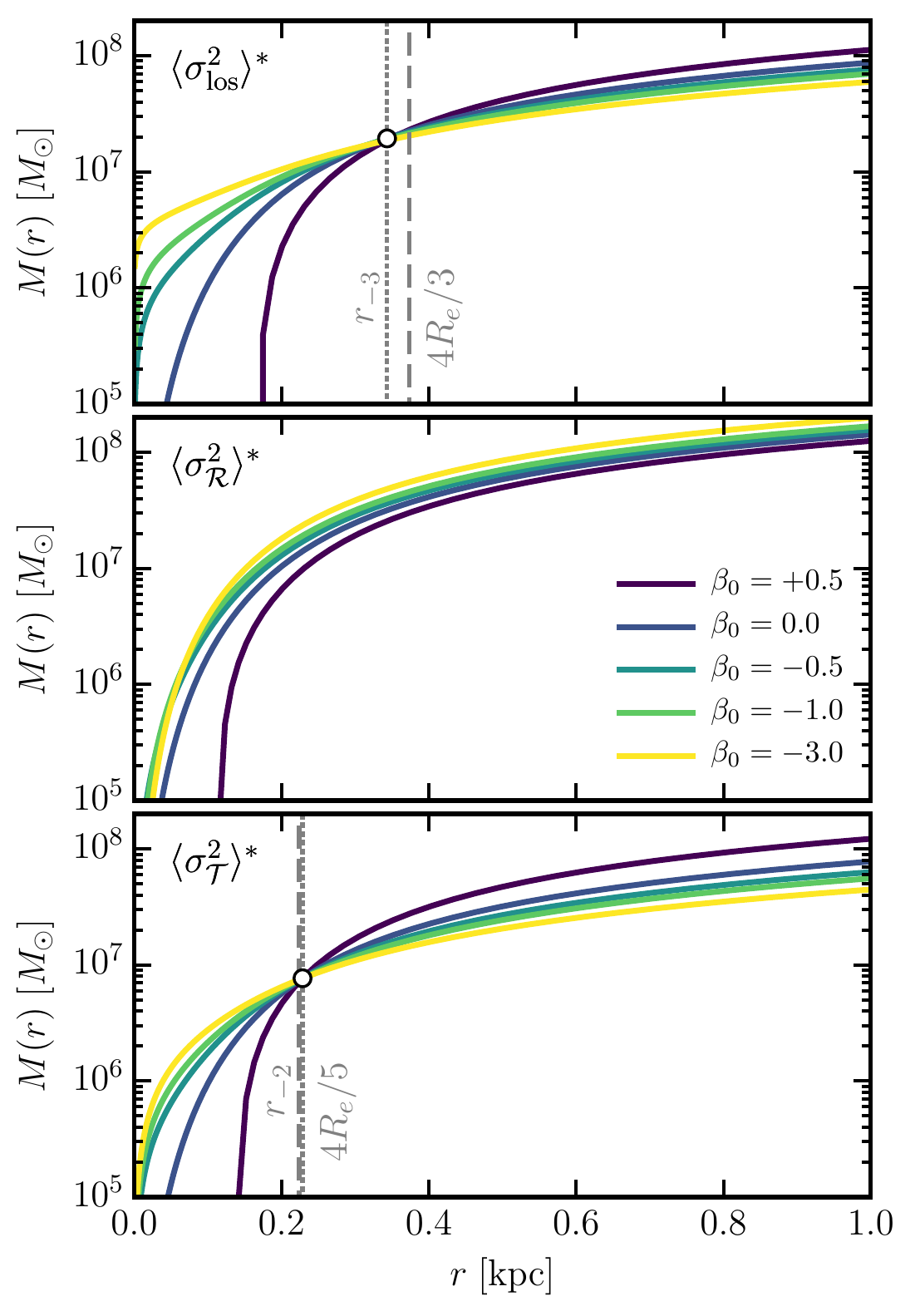}
    \caption{---
        {\bf \emph{Radii of minimized uncertainty for idealized models}}. 
        Curves depict the cumulative mass profiles derived in Appendix~\ref{sec:mass.profiles} based on fixed line-of-sight (top panel), parallel (middle panel), and transverse (bottom panel) velocity dispersions, all of which use the median values for Sculptor from Table~\ref{tab:1}. The results assume that the observed dispersion profile for each component is constant with $R$. The lines correspond to several choices of constant intrinsic anisotropy $\beta(r) = \beta_{0}$ as indicated by the colors.
        We also assume that the underlying tracer profile follows a Plummer model. The small white circles in the top plot and bottom plot show masses predicted by the $M_{-3}$ and $M_{-2}$ estimators, respectively.
        These points intersect the region of mass that is independent of the anisotropy.  Note that the parallel component has no such intersection, as anticipated in Section~\ref{sec:parallel}. The dotted lines give the characteristic log-slope radii of the tracer profile while the dashed shows the standard mapping to the projected observable, $R_{e}$. 
    }
    \label{fig:1}
\end{figure}

\subsection{\emph{Plane of the sky}: Measurements Parallel to \emph{R}}
\label{sec:parallel}
Consider a dispersion-supported stellar system that is well studied, such that $\sigma_{\mathcal{R}}(R)$ is determined accurately through observations. We begin by relating the projected measurement of $\sigma_{\mathcal{R}}(R)$ to the intrinsic quantities via Eq.~\eqref{eq:sig_projR.mapping}. This is then rewritten to an invertable form (see Appendix~\ref{sec:mass.profiles}),
\begin{align}
    \Sigma_{\star} \sigma_{\mathcal{R}}^{2}(R)
    &=
    \int_{R^{2}}^{\infty}
    \frac{dr^{2}}{\sqrt{r^{2} - R^{2}}}
    \left[
    n_{\star}\sigma_{r}^{2}
    -
    \int_{r^{2}}^{\infty} d\tilde{r}^{2} \frac{\beta n_{\star} \sigma_{r}^{2}}{2\tilde{r}^{2}}
    \right] \, .
    \label{eq:sig_projR.mapping.inv}
\end{align}
From its invertable form, the left-hand side is an accurate, observable quantity that is ignorant of the form of $\beta$. Therefore, the term in the brackets must be a well-defined quantity regardless of the form of $\beta$ chosen. Therefore, we are allowed to consider the simple case of equating the isotropic integrand with an integrand that is dependent on some arbitrary anisotropy:
\begin{align}
    n_{\star}\sigma_{r}^{2}\big|_{\beta = 0} 
    &= 
    n_{\star}\sigma_{r}^{2}  - \int_{r^{2}}^{\infty} d\tilde{r}^{2} \frac{\beta n_{\star} \sigma_{r}^{2}}{2\tilde{r}^{2}}
    \, .
\end{align}
By then taking the derivative in respect to $\log r$ and introducing a factor of $r\sigma_{r}^{2}/G$ to both sides, the left-hand and right-hand side are allowed to be rewritten in the form of the integrated Jeans masses, allowing use to express the difference:
\begin{align}
    M(r|\beta) - M(r|0)
    &=
    \frac{r \sigma_{r}^{2} \beta}{G} 
    \, .
    \label{eq:parallel_mass_diff}
\end{align}
Importantly, this expression lacks the parenthetical term seen in Eq.~\eqref{eq:los_mass_diff}. We conclude that a radius that minimizes the anisotropy, at least, for the assumptions we considered in the Jeans modeled measurements of $\sigma_{\mathcal{R}}(R)$, {\it does not} exist in whatever limiting case of $\beta$ we were to impose, since the anisotropy is a dependent quantity throughout the mass profile.

\subsection{\emph{Plane of the sky}: Measurements Transverse to \emph{R}}
\label{sec:tangential}
Consider a dispersion-supported stellar system that is well studied, such that $\sigma_{\mathcal{T}}(R)$ is determined accurately through observations, in which all stars are bounded inside the system. We begin by relating of $\sigma_{\mathcal{T}}(R)$ to $\sigma_{r}(r)$, given by Eq.~\eqref{eq:sig_projT.mapping}. Fortunately, this is already in an invertable form; we now equate its isotropic and general anisotropic form to one another 
\begin{align}
    n_{\star}\sigma_{r}^{2}\big|_{\beta = 0} 
    &= 
    n_{\star}\sigma_{r}^{2}(1-\beta)
    \, ,
\end{align}
differentiate it with respect to $\log r$, and algebraically manipulate to acquire the expression
\begin{align}
    M(r|\beta) - M(r|0)
    &=
    \frac{r \sigma_{r}^{2} \beta}{G} 
    \left(\gamma_{\star} + \gamma_{\sigma} + \gamma_{\beta} - 2 \right)
    \, .
    \label{eq:projT_mass_diff}
\end{align}
We see that {\it there can} exist a radius, that we shall denote as $\tilde{r}_{\mathrm{eq}}$,\footnote{This is not to be associated with the radius, $r_{\rm eq}$, seen in the derivation of $M_{-3}$, as that $r_{\rm eq}$ is constrained to measurements of $\sigma_{\rm los}$. Simply, the $r_{\rm eq}$ of Eq.~\eqref{eq:wolf_constraint} and $\tilde{r}_{\rm eq}$ of Eq.~\eqref{eq:lazar_constraint} are taken to be nonequivalent.} where the parenthesis vanishes. The possible existence of $\tilde{r}_{\mathrm{eq}}$ therefore minimizes the dependency of $\beta$ around the region $\tilde{r}_{\mathrm{eq}}$ for measurements solely based off of $\sigma_{\mathcal{T}}(R)$, such that,
\begin{align}
    \gamma_{\star}(\tilde{r}_{\mathrm{eq}}) 
    &= 
    2 - \gamma_{\sigma}(\tilde{r}_{\mathrm{eq}}) - \gamma_{\beta}(\tilde{r}_{\mathrm{eq}})
    \, .
    \label{eq:lazar_constraint}
\end{align}
Moreover unless galaxies have large variation in $\sigma_{r}^{2}$ and in $\beta$ with radius, we may expect $\gamma_\sigma(\tilde{r}_{\rm eq}) + \gamma_\beta(\tilde{r}_{\rm eq})  \ll 2$, as least for radii in the vicinity of $\tilde{r}_{\rm eq} < r_{-3} \simeq r_{1/2}$ for commonly assumed stellar density profiles. Therefore, we can expect that to a good approximation, $\tilde{r}_{\rm eq} \simeq r_{-2}$, where $r_{-2}$ is the radius at which the log-slope of the tracer profile is equivalent to $-2$.

Like before, we now consider the integrated Jeans mass. The dependence of $\beta$ can be absorbed into the definition of the intrinsic total velocity dispersion. Moreover, the formulation of Eq.~\eqref{eq:total.dispersion} allows,
$(1-\beta)\sigma_{r}^{2} = \sigma_{\theta}^{2} = \sigma_{\mathcal{T}}^{2}$ with the assumption of spherical symmetry. The Jeans equation becomes
\begin{align}
    \frac{GM(r)}{r}
    &=
    2\sigma_{\theta}^2(r)
    + (\gamma_{\star} - \gamma_{\sigma} - 2)\sigma_{r}^{2}(r) \, .
\end{align}
If in fact that $\gamma_\star + \gamma_\sigma \approx \gamma_\star \simeq 2$, the term in parenthesis vanishes at the radius $r_{-2}$. The remaining term on the right-hand side depends only on the intrinsic transverse component, $\sigma_{\theta}$ = $\sigma_\mathcal{T}$, which is an observable.\footnote{Note that the term in brackets in Eq.~\eqref{eq:sig_projT.mapping} is constrained by observables. Specifically, the intrinsic transverse dispersion, $\sigma_{\theta}$, it is related to the transverse component along the plane of the sky via $(1-\beta)\sigma_{r}^{2} = \sigma_{\theta}^{2} = \sigma_{\mathcal{T}}^{2}$. This is what allows proper motion mesurements to constrain the anisotropy \citep{strigari2007astrometry}.}
Finally, we obtain an idealized estimator
\begin{align}
    M_{-2}^{\rm ideal} &\equiv M(r_{-2})
    =
    \frac{2 \langle \sigma_{\mathcal{T}}^{2} \rangle^{*} r_{-2}}{G}
    \, ,
    \label{eq:lazar_mass}
\end{align}
where we have assumed $\sigma_{\theta}(r_{-2}) \simeq \sigma_{\mathcal{T}}^{2}(r_{-2}) \simeq \langle \sigma_{\mathcal{T}}^{2} \rangle^{*}$. The implied circular velocity at $r_{-2}$ is particularly succinct
\begin{align}
    V_{\rm circ}(r_{-2})
    &=
    \sqrt{\ 2 \langle \sigma_{\mathcal{T}}^{2} \rangle^{*}}
    \label{eq:lazar_vcirc}
    \, .
\end{align}
From here-on, we will refer Eq.~\eqref{eq:lazar_mass} as $M_{-2}$ and Eq.~\eqref{eq:lazar_vcirc} as $V_{-2}$.

\begin{figure}
    \centering
    \includegraphics[width=\columnwidth]{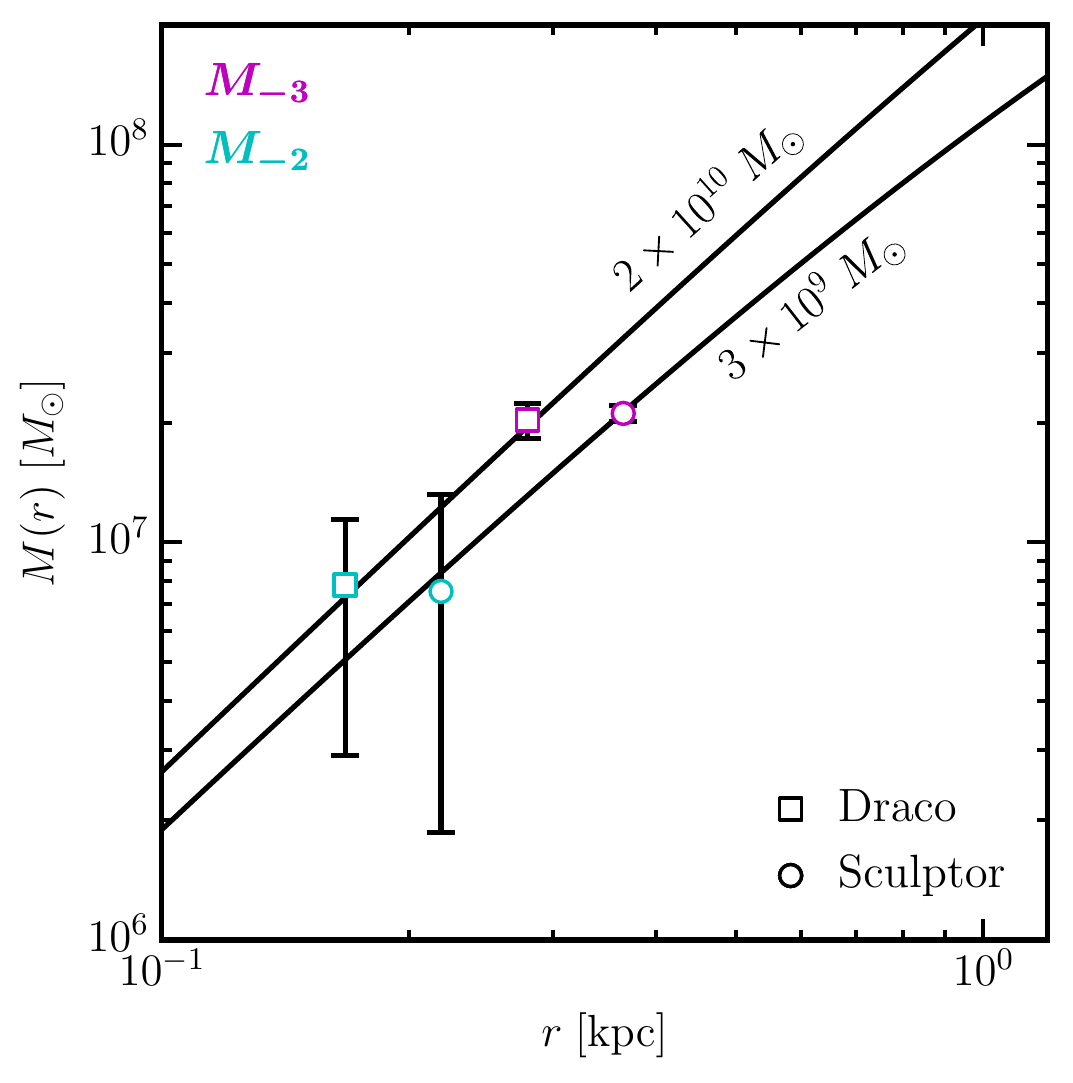}
    \caption{---
        {\bf \emph{Mass measurements for Draco and Sculptor}}. 
        For each galaxy, points correspond to the line-of-sight mass (magenta), $M_{-3}$, and the projected tangential mass (cyan), $M_{-2}$, at two characteristic radii. Lines show representative NFW mass profiles of fixed $M_{\rm vir}$ with median concentration set by subhalos in the {\tt Phat-ELVIS} simulations. 
    }
    \label{fig:2}
\end{figure}

\subsection{Overview of Assumptions}
\label{sec:assumptions}
We have made a few assumptions in the derivation of $M_{-3}$ and $M_{-2}$. In addition to the strong assumption that galaxies are spherical,  we have assumed that the velocity dispersions are relatively flat such that $\sigma_{\rm tot}^{2}(r_{-3}) \simeq \langle \sigma_{\rm tot}^{2} \rangle^{*} = 3\langle \sigma_{\rm los}^{2} \rangle^{*}$ and $\sigma_{\rm \theta}^{2}(r_{-2}) = \sigma_{\mathcal{T}}^{2}(r_{-2}) \simeq \langle \sigma_{\mathcal{T}}^{2} \rangle^{*}$.  \cite{wolf2010accurate} showed that the assumption for the line-of-sight component is excellent for a variety of models that match line-of-sight data, yet, for the transverse component, not enough data is available to test this assumption. Some justification comes from Section~\ref{sec:mock.observations}, where we use a set of cosmological simulations of dwarf galaxies in mock observations and find that these assumptions are good to better than $10\%$. 

Second, we have assumed that the intrinsic radial velocity dispersion varies minimally with radius compared to the tracer profile out to $r_{-3}$. More specifically we assume that the log-slopes of the tracer velocity dispersion profiles are small compared to $-3$ and $-2$ at $r_{-3}$ and $r_{-2}$, respectively. Third, we assume that the velocity dispersion anisotropy varies slowly with radius compared to the light profile.  If $\beta(r)$ and $\sigma_{r}(r)$ vary quickly as a function of radius $r$, then the mass estimators will break down. 

\begin{figure}
    \centering
    \includegraphics[width=\columnwidth]{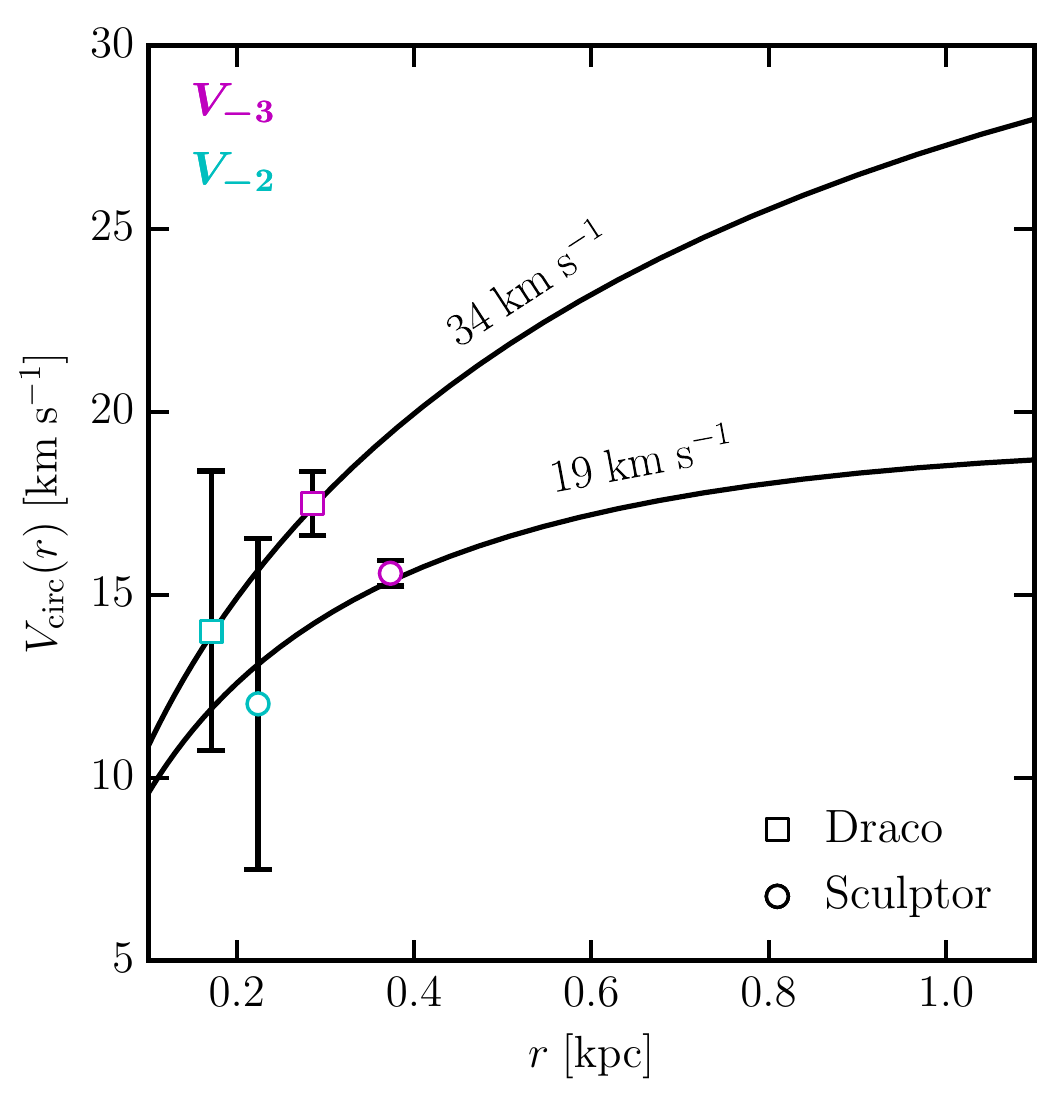}
    \caption{---
        {\bf \emph{Observed circular velocities of Draco and Sculptor}}. 
        Circular velocity curves for NFW subhalos of a given $V_{\rm max}$ are shown for the two characteristic radii. Each assumes a median $r_{\rm max}$ as derived from the {\tt Phat-ELVIS} simulations.
    }
    \label{fig:3}
\end{figure}

In order to map $M_{-3}$ and $M_{-2}$ to observables measured in two-dimensions, the characteristic radii of the tracer profile, $r_{-2}$ and $r_{-3}$, must be mapped to the projected tracer profile that is observed. If we assume that the three-dimensional profile is well described by a \cite{plummer1911problem} profile, then $r_{-3} \simeq 4R_{e}/3$ and $r_{-2} \simeq 4R_e/5$. That is
\begin{align}
    M(3r_{1/2}/5)
    &\simeq
    \frac{6 \langle \sigma_{\mathcal{T}}^{2}\rangle^{*} r_{1/2}}{5G};
    \qquad
    M(4R_{e}/5)
    \simeq
    \frac{8 \langle \sigma_{\mathcal{T}}^{2}\rangle^{*} R_{e}}{5G}
    \, .
\end{align}
To clarify, {\em if the underlying three-dimensional tracer profile is not well-described by a Plummer profile, then this mapping will fonder}. Ultimately the mapping between the three-dimensional characteristic radii and observed two-dimensional radii will obey another relationship that depends on the underlying profile. 

Fig.~\ref{fig:1} provides a test and illusration of the derivation presented above using a full mass profile analysis derived in Appendix~\ref{sec:mass.profiles}. Shown are the mass profiles implied by various choices of constant velocity dispersion anisotropy constrained by dispersion components along the line-of-sight (Eq.~\ref{eq:los.mass.profile.constant}; top panel), parallel (Eq.~\ref{eq:projR.mass.profile.constant}; middle panel), and tangential (Eq.~\ref{eq:projT.mass.profile.constant}; bottom panel) under the assumption of constant $\beta$ (denoted by $\beta_{0}$). We assume that the velocity dispersions for each component is constant with $R$, and set them equal to the luminosity-weighted median values observed for Sculptor (9, 11.5, and 8.5 km s$^{-1}$, respectively). We also assume that the tracer profile follows a Plummer, again matched to the median value for Sculptor given in Table~\ref{tab:1}. The white circles show the estimators $M_{-3}$ and $M_{-2}$  in the top and bottom panels, respectively. Encouragingly, they intersect the regions where all of the varying $\beta_{0}$ mass profiles converge. As anticipated in Section~\ref{sec:parallel}, constraints imposed by the parallel component, $\sigma_{\mathcal{R}}$, show no convergence point. This figure shows that the mass estimators we have derived work under reasonable, but idealized assumptions. In the last last section of Appendix~\ref{sec:mass.profiles}, we show a similar analysis that allows for parametric forms of $\beta(r)$ commonly used in Jeans modeling analyses.  We show the idealized mass estimators work well unless $\beta(r)$ varies rapidly with radius (as expected).  

Of course, real galaxies will not obey these assumptions with absolute precision. Perhaps most importantly, no galaxy is perfectly spherical. We expect real galaxies to have velocity dispersion profiles that vary with radius to some degree. Galaxies also have light tracer profiles that will not necessarily obey convenient functional characterizations such as the Plummer model in three-dimensions, which will make determining $r_{-2}$ and $r_{-3}$ more difficult.  We test these assumptions along with our estimator in Section \ref{sec:mock.observations} using cosmological dwarf galaxy simulations.

\section{Modeling from Observations}
\label{sec:observation.results}
We now apply our mass estimator using kinematic measurements for the spheroidal galaxies, Draco and Sculptor. Table~\ref{tab:1} lists the observed properties that we adopt. We assume that each galaxies stellar distribution obeys a \cite{plummer1911problem} profile in deprojection and in projection. We used the radial conversions for a Plummer profile given in \cite{wolf2010accurate}.

\subsection{The Internal Structure of Draco and Sculptor}
Fig.~\ref{fig:2} plots the implied mass of Draco (squares) and Sculptor (circles) using both $M_{-3}$ (magenta colored) and $M_{-2}$ (cyan colored). With the current data today, masses implied from well studied, line-of-sight measurements have smaller error bars while the implied masses from the tangential along the plane of the sky have relatively larger error bars. Also plotted are the NFW \citep{navarro1997universal} mass profiles at fixed halo mass, $M_{\rm vir}= 2\times 10^{10}$ and $3\times 10^{9}\ M_{\odot}$. Concentrations are set to $16.3$ and $20.2$, respectively, based on the median values for subhalos of this mass in the $z=0$ dark matter only physics results of the {\tt Phat-ELVIS} simulations \citep{kelley2019phat}. The subhalo masses were chosen so that at median value of the concentration for the profiles intersect the line-of-sight mass points. In principle, by comparing the location of the tangentially-derived masses to the extrapolated NFW curves can allow us to determine if the predictions are consistent with a cuspy profile. Both galaxies appear consistent with sitting within a typical CDM halo. Note that this result for Draco is in agreement with results by \cite{read2018case}, who find Draco to be cusped around the same radial range. 


\renewcommand{\arraystretch}{1.0}%

\begin{table}
    \centering
    \caption{ 
        Global properties of the $10^{10}\ M_{\odot}$ galaxies at $z=0$. 
        }
    \begin{threeparttable}
    \begin{tabular}{cccccc} 
        \toprule
        {Halo} & 
        {$M_{\rm vir}$} & {$M_{\star}$} &
        {$r_{1/2}$} &
        {$\rm DM$} &
        {References}
        \\
        {Name} & 
        {$[M_{\odot}]$} & {$[M_{\odot}]$} & 
        {$[\rm pc]$} &
        {Core?} &
        {}
        \\ 
        \midrule
        & 
        {(1)} & {(2)} & 
        {(3)} & 
        {(4)} &
        {}
        \\
        \midrule 
        \multicolumn{5}{l}{\bf Cold\ Dark\ Matter} \\
        \midrule
        {m10b} & 
        {$9.29 \times 10^{9}$} & {$4.65 \times 10^{5}$} &
        {340} & 
        {\xmark} &
        {$a,b,c$}
        \\
        {m10c} & 
        {$8.92 \times 10^{9}$} & {$5.75 \times 10^{5}$} &
        {350} & 
        {\xmark} &
        {$a,b,c$}
        \\
        {m10d} & 
        {$8.43 \times 10^{9}$} & {$1.53 \times 10^{6}$} &
        {530} &
        {\xmark} &
        {$a,c$}
        \\
        {m10e} & 
        {$1.02 \times 10^{10}$} & {$1.98 \times 10^{6}$} &
        {620} &
        {\xmark} &
        {$a,c$}
        \\
        \midrule 
        \multicolumn{5}{l}{\bf Self-interacting\ Dark\ Matter;\ $\sigma/m = 1\ \rm cm^{2}\ g^{-1}$}
        \\
        \midrule
        {m10b} & 
        {$8.13 \times 10^{9}$} & {$1.05 \times 10^{6}$} &
        {504} &
        {\cmark} &
        {$b,c$}
        \\
        {m10c} & 
        {$8.71 \times 10^{9}$} & {$7.48\times 10^{5}$} &
        {430} &
        {\cmark} &
        {$c$}
        \\
        {m10d} & 
        {$8.10 \times 10^{9}$} & {$1.37 \times 10^{6}$}  &
        {591} & 
        {\cmark} &
        {$b,c$}
        \\
        {m10e} & 
        {$9.95 \times 10^{9}$} & {$1.63 \times 10^{6}$}  &
        {572} & 
        {\cmark} &
        {$c$}
        \\
        \bottomrule
    \end{tabular}
    \end{threeparttable}
    \begin{tablenotes}
        \item[](1): 
        The mass of the dark matter halo defined by the background virial overdensity \protect\citep{bryan1998statistical}.
        \item[](2): 
        The stellar mass of the galaxy; $M_{\star} := M_{\rm vir}(< 10\%\ r_{\rm vir})$.
        \item[](3): 
        The deprojected radius that contains half of $M_{\rm star}$.
        \item[](4):  
        Verification that a dark matter core has formed.
        \item[]
        {\bf \emph{References}} --
        $(a)$: \protect\cite{fitts2017fire},
        $(b)$: \protect\cite{robles2017fire},
        $(c)$: \protect\cite{fitts2019baryons}.
    \end{tablenotes}
    \label{tab:2}
\end{table}


Fig.~\ref{fig:3} provides an alternative view by plotting observed circular velocities using now $V_{-3}$ and $V_{-2}$. The rotation curves for NFW profiles at fixed values of $V_{\rm max}=$ 19 and $34\ \rm km\ s^{-1}$ are also plotted, with median values of $r_{\rm max} = 1.67$ and $4.71$ kpc, respectively, for the same subhalos of {\tt Phat-ELVIS}. As seen previously in Fig.~\ref{fig:2}, both measurements are consistent with the expectations for an NFW. Sculptor's median does fall below the extrapolated NFW, though it is easily consistent within error. If Sculptor has a cored inner-density it could have interesting implications. With a stellar mass of $M_{\star} \simeq 4\times 10^{6}\ M_{\odot}$, this galaxy lies near the low-mass edge of where feedback may be able to produce significant cores \citep{bullock2017small}. This motivates the acquisition of additional data to provide a more precise measure of $\langle \sigma_{\mathcal{T}}^{2} \rangle^{*}$.

\section{Mock Observations}
\label{sec:mock.observations}
We are now interested in testing the mass estimators discuss previously, including the one derived here for the first time. We use simulations that have been previously published with data kindly supplied by the authors \citep{fitts2017fire,robles2017fire}.  The simulations were run as part of the
Feedback in Realistic Environments (\texttt{FIRE}) project and include galaxies simulated in both Cold Dark Matter (CDM) and Self Interacting Dark Matter (SIDM). 
Table \ref{tab:2} lists the global parameters of the galaxies considered herein as well as the references the reader can refer to with the specific physics used when running the {\tt FIRE-2} algorithm \citep{hopkins2014galaxies,hopkins2015new,hopkins2018fire}.

We specifically have chosen low-mass galaxies that are dispersion supported that resemble dwarf spheroidals. The values of $M_{\star}/M_{\rm vir}$ for the CDM galaxies do not produce enough energy to transform cusps to cores and thus provide a good test for ``cuspy" underlying profiles \citep{di2013dependence,chan2015impact,tollet2016nihao,bose2019bursty}, while SIDM halos are naturally core-like. Their stellar masses are low enough that episodic gas outflows do not  bias estimates from equilibrium when using Jeans modeling. \citep{elbardy2016breathing,elbadry2017jeans}. In summary, we consider two types of simulations:

\begin{figure*}
    \centering
    \includegraphics[width=\textwidth]{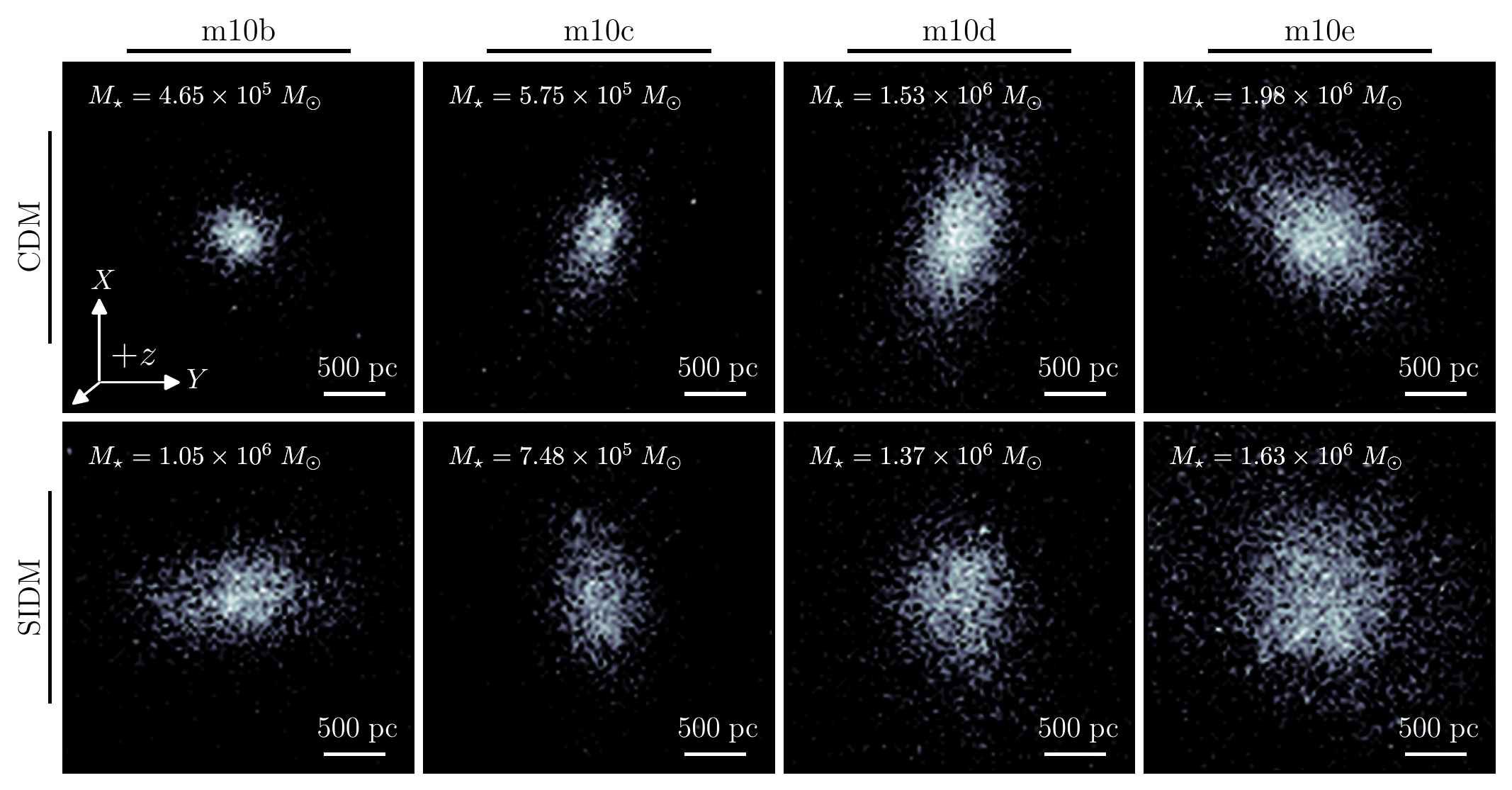}
    \caption{---
    {\bf \emph{Mock observations of galaxies in the plane of the sky}}: The stellar surface density of the stars for our dwarf galaxies (given in columns) actualized along a random orientation of the plane ($X,Y$) looking along the line-of-sight, $z$, in CDM (top row) and SIDM (bottom row). 
    The center of mass of the galaxy determined from this plane of observation is centered at the origin. This shows how elongated several galaxies can appear projections when viewed in the plane of the sky for observations.
    }
    \label{fig:4}
\end{figure*}

\begin{itemize}
    \item[] {\bf \emph{CDM}}:
    Dark matter is considered to be collisionless. 
    The sample of galaxies simulated in CDM are m10b, m10c, m10d, and m10e, which were first presented in \cite{fitts2017fire} and explored further in \cite{fitts2018assembly,fitts2019baryons}.
    The fiducial CDM simulations have a baryonic particle mass of $m_{\rm b} = 500\ M_{\odot}$ with force resolution $\epsilon_{b} = 2\ \rm pc$ and a dark matter mass $m_{\rm DM} = 2500\ M_{\odot}$ with softening $\epsilon_{\rm DM} = 35\ \rm pc$. This sample of galaxies have their dark halos forming cusps $z=0$.
    \vspace{1ex}
    
    \item[] {\bf \emph{SIDM}}:
    This considers the CDM power spectrum but with a imposed self-interaction cross section of $\sigma/m = 1\ \rm cm^{2}\ g^{-1}$ that is velocity independent.  The sample of galaxies considered here are the analogs of the CDM galaxies: m10b, m10c, m10d, and m10e. SIDM analogs of m10d and m10b were first presented in \cite{robles2017fire} and further explored with m10c and m10e in \cite{fitts2019baryons}. A key result is that all halos have formed appreciable cores at $z=0$.
\end{itemize}

\subsection{Methodology}
\subsubsection{Properties in three-dimensions}
The center position of the galaxies are determined by using an iterative "shrinking spheres" method \citep{power2003inner,navarro2004inner}. That is, the center of mass of star particles is successively computed in a sphere and then has its radius reduced by $50\%$, which is then re-centered on the new center of mass. This is done iteratively until a thousand particles enclose the minimized sphere. The center of mass velocity is then computed using all of the star particles enclosing the final minimized radius. Three-dimensional positions and velocities of all the star particles, associated with that galaxy, are then transformed to be relative to the center of mass position and velocity, respectively.

The stellar profiles are assembled using 25, log-spaced radial  bins of starting from the center of mass of the stars out to $4 \times r_{1/2}$. In quantifying the characteristic radii of $r_{-2}$ and $r_{-3}$, the stellar profiles are smoothed using a third ordered spline fit as profiles tend to be noisy. From there, $r_{-2}$ and $r_{-3}$ are interpolated from the log-gradients of the resultant fits. In the construction of the intrinsic dispersion profiles, the Cartesian velocities relative to the center of mass are transformed to spherical coordinates and are evaluate using the same spherical bin spacing. In each bin shell of $r$, the relative velocities are weighted by their associated stellar particle mass. This includes both the random motions and streaming motions.

We also compared between a sample containing {\it only} star particles bound to the dynamical system and another sample containing {\it both} bound and unbound star particles to the dynamical system. Results for these two population samples were found to be indistinguishable, as unbound star particles only comprised $1\%$ of the galaxies' stellar population. Final results presented here include both bound stars and unbound stars. 

\subsubsection{Idealized Mock Observations}
For each galaxy, we construct 1000 mock observations. That is, mock observations are done in 1000 random orientations with each orientation evaluated as follows: the relative Cartesian positions and velocities of the galaxies' stellar particles are rotated into a new orientation denoted by prime coordinates, such that the star particles along the new line-of-sight axis, $z'$ with velocity $v_{z'} \equiv v_{\rm los}$, are stacked along the projected $x'-y'$ plane. From the galaxy projected on this plane, the center position is determined by re-implementing an iterative "shrinking spheres" method. This again determines the center-of-mass position and velocities of the stars found on the $x'-y'$ plane. We define this as the center of the galaxy when analysing its projection in two-dimensions, where we now label the center position and velocity as $\boldsymbol{X} = (X,Y)$ and $\boldsymbol{V} = (V_{X},V_{Y})$, respectively. Hereafter, we drop the prime notation for the line-of-sight axis.

Fig.~\ref{fig:4} illustrates a single mock observations by projecting the stellar particles of each galaxy using the method discussed in the previous paragraph. These images have been made after the transformation of coordinates and placing the origin at the center of mass from the projected distribution of stars. Note that for both CDM and SIDM the galaxies are not spherical but do appear to have morphologies comparable to actual observed dwarf spheroidals. That is, dwarf galaxies can appear elongated in the plane of sky (plane $X-Y$ in the figure).  


\setlength{\tabcolsep}{4.25pt}
\renewcommand{\arraystretch}{1.15}%

\begin{table*}
    \centering
    \caption{ 
        Simulated properties of our galaxies in relation to the assumptions summarized at the end of Sec.~\ref{sec:estimators}. Uncertanties are quoted as the 1$\sigma$ dispersion from the median. Values measured based off of the characteristic radii are given by columns (1-8) and measured values using only $R_{e}$ are given in columns (9-12).
         }
    \begin{threeparttable}
    \begin{tabular}{c|cc|cc|cc|cc|cc|cc} 
        \toprule
        {Halo} & 
        {$\gamma_{\sigma}(r_{-2})$} &
        {$\gamma_{\sigma}(r_{-3})$} &
        {$\widetilde{\sigma}_{\mathcal{T},-2} $} &
        {$\widetilde{\sigma}_{\rm tot,-3}$} &
        {$V_{-2}/V_{\rm true}$} & 
        {$V_{-3}/V_{\rm true}$} &
        {$\xi_{\rm est}$} &
        {$\xi_{\rm true}$} &
        {$V_{-2}^\prime/V_{\rm true}$} & {$V_{-3}^\prime/V_{\rm true}$} & 
        {$\widetilde{R}_{e,-2}$} & 
        {$\widetilde{R}_{e,-3}$} 
        \\
        \midrule
        & 
        {(1)} & 
        {(2)} & 
        {(3)} & 
        {(4)} &
        {(5)} & 
        {(6)} &
        {(7)} & 
        {(8)} &
        {(9)} & 
        {(10)} &
        {(11)} & 
        {(12)} 
        \\
        \midrule 
        \multicolumn{5}{l}{{\bf Cold\ Dark\ Matter};\ (Cusps)} \\
        \midrule
        {m10b} & 
        {$0.11$} & 
        {$0.16$} &
        {$0.96^{+0.04}_{-0.04}$} & 
        {$0.98^{+0.03}_{-0.03}$} &
        {$0.96^{+0.01}_{-0.01}$} & 
        {$0.92^{+0.03}_{- 0.03}$} &
        {$0.45^{+0.08}_{-0.08}$} &
        {$0.54$} &
        {$1.09^{+0.13}_{-0.12}$} &
        {$1.02^{+0.14}_{-0.14}$} & 
        {$0.83^{+0.11}_{-0.10}$} & 
        {$0.80^{+0.11}_{-0.10}$}
        \\
        {m10c} & 
        {$0.09$} & 
        {$0.74$} &
        {$0.97^{+0.04}_{-0.04}$} & 
        {$1.02^{+0.06}_{-0.04}$} &
        {$0.99^{+0.01}_{-0.01}$} & 
        {$0.96^{+0.04}_{-0.05}$} &
        {$0.51^{+0.12}_{-0.10}$} &
        {$0.57$} &
        {$1.10^{+0.05}_{-0.06}$} &
        {$1.05^{+0.08}_{-0.09}$} & 
        {$0.86^{+0.04}_{-0.05}$} & 
        {$0.84^{+0.04}_{-0.04}$} 
        \\
        {m10d} & 
        {$0.02$} & 
        {$0.56$} &
        {$1.01^{+0.04}_{-0.04}$} & 
        {$0.95^{+0.04}_{-0.04}$} &
        {$1.00^{+0.02}_{-0.02}$} & 
        {$1.01^{+0.05}_{-0.04}$} &
        {$0.47^{+0.11}_{-0.11}$} &
        {$0.46$} &
        {$1.00^{+0.04}_{-0.04}$} & 
        {$1.01^{+0.07}_{-0.07}$} &
        {$1.01^{+0.03}_{-0.03}$} & 
        {$0.98^{+0.03}_{-0.03}$}
        \\
        {m10e} & 
        {$0.21$} & 
        {$0.48$} &
        {$0.97^{+0.04}_{-0.03}$} & 
        {$0.93^{+0.06}_{-0.06}$} &
        {$1.01^{+0.04}_{-0.04}$} & 
        {$1.02^{+0.07}_{-0.07}$} &
        {$0.64^{+0.19}_{-0.19}$} &
        {$0.63$} &
        {$1.09^{+0.11}_{-0.09}$} & 
        {$1.12^{+0.12}_{-0.12}$} &
        {$0.89^{+0.06}_{-0.05}$} & 
        {$0.86^{+0.06}_{-0.05}$}
        \\
        \midrule 
        \multicolumn{5}{l}{{\bf Self-Interacting\ Dark\ Matter};\ (Cores)}
        \\
        \midrule
        {m10b} & 
        {$0.34$} & 
        {$0.44$} &
        {$1.01^{+0.05}_{-0.06}$} & 
        {$1.04^{+0.08}_{-0.06}$} &
        {$1.02^{+0.04}_{-0.04}$} & 
        {$0.93^{+0.04}_{-0.04}$} &
        {$0.46^{+0.08}_{-0.08}$} &
        {$0.75$} &
        {$1.14^{+0.09}_{-0.10}$} & 
        {$1.04^{+0.12}_{-0.12}$} &
        {$0.87^{+0.05}_{-0.05}$} & 
        {$0.84^{+0.05}_{-0.04}$}
        \\
        {m10c} & 
        {$0.20$} & 
        {$0.26$} &
        {$0.96^{+0.05}_{-0.05}$} & 
        {$0.91^{+0.08}_{-0.06}$} &
        {$1.15^{+0.04}_{-0.06}$} & 
        {$1.02^{+0.09}_{-0.08}$} &
        {$0.50^{+0.12}_{-0.10}$} &
        {$0.84$} &
        {$1.36^{+0.09}_{-0.10}$} & 
        {$1.25^{+0.13}_{-0.14}$} &
        {$0.82^{+0.04}_{-0.04}$} &  
        {$0.79^{+0.04}_{-0.04}$}
        \\
        {m10d} & 
        {$0.50$} & 
        {$0.76$} &
        {$0.94^{+0.03}_{-0.03}$} & 
        {$1.13^{+0.03}_{-0.04}$} &
        {$1.08^{+0.02}_{-0.02}$} & 
        {$0.93^{+0.03}_{-0.03}$} &
        {$0.46^{+0.11}_{-0.11}$} &
        {$0.77$} &
        {$1.27^{+0.05}_{-0.05}$} & 
        {$1.18^{+0.06}_{-0.05}$} &
        {$0.82^{+0.02}_{-0.02}$} & 
        {$0.79^{+0.02}_{-0.02}$}
        \\
        {m10e} & 
        {$0.17$} & 
        {$0.49$} &
        {$0.97^{+0.03}_{-0.03}$} & 
        {$1.02^{+0.08}_{-0.08}$} &
        {$0.99^{+0.05}_{-0.03}$} & 
        {$1.03^{+0.07}_{-0.08}$} &
        {$0.64^{+0.19}_{-0.19}$} &
        {$0.62$} &
        {$1.02^{+0.11}_{-0.08}$} & 
        {$1.06^{+0.14}_{-0.13}$} &
        {$0.97^{+0.05}_{-0.05}$} & 
        {$0.93^{+0.04}_{-0.05}$}
        \\
        \bottomrule
    \end{tabular}
    \end{threeparttable}
    \begin{tablenotes}
        \item[](1): 
        The value of the log-slope of the radial dispersion profile at $r_{-2}$.
        \item[](2): 
        The value of the log-slope of the radial dispersion profile at $r_{-3}$.
        \item[](3): 
        The ratio of the stellar transverse velocity dispersion at $r_{-2}$ normalized by the weighted median measurement: 
        \\
        $\widetilde{\sigma}_{\mathcal{T},-2} := \sigma_{\mathcal{T}}(r_{-2})/\langle \sigma_{\mathcal{T}}\rangle^{*}$.
        \item[](4):  
        The ratio of the stellar total velocity dispersion at $r_{-3}$ normalized by the weighted line-of-sight measurement:
        \\
        $\widetilde{\sigma}_{\rm tot,-3} := \sigma_{\rm tot}(r_{-3})/(\sqrt{3}\langle \sigma_{\rm los}\rangle^{*})$.
        \item[](5):  
        Ratio between the value of $V_{-2}$, Eq.~\protect\eqref{eq:lazar_vcirc} at $r_{-2}$, to the true dynamical circular velocity, $V_{\rm circ}$.
        \item[](6):
        Ratio between the value of $V_{-3}$, Eq.~\protect\eqref{eq:wolf_vcirc} at $r_{-3}$, to the true value of $V_{\rm circ}$. 
        \item[](7):
        The implied power-law slope of the circular velocity profile using the mass estimators at the two characteristic radii, $\xi_{\rm est} := \Delta \log V_{\rm est}/\Delta \log r $.
        \item[](8):
        The true slope of the dynamical component of the circular velocity curve at these two characteristic radii assuming a power-law.
        \item[](9):
        Ratio between the estimator value of $V$ that approximates $r_{-2} = 4R_{e}/5$ from a forced Plummer fit, to the true value of $V_{\rm circ}$.
        \item[](10):
        Ratio between the estimator value of $V$ that approximates $r_{-3} = 4R_{e}/3$ from a forced Plummer fit, to the true value of $V_{\rm circ}$.
        \item[](11):
        Ratio between the value of $r_{-2}$ and fitted value of $R_{e}$ forcing a Plummer profile: $\widetilde{R}_{e,-2} := (4R_{e}/5)/r_{-2}$.
        \item[](12):
        Ratio between the value of $r_{-3}$ and fitted value of $R_{e}$ forcing a Plummer profile:$\widetilde{R}_{e,-3} := (4R_{e}/3)/r_{-3}$.
    \end{tablenotes}
    \label{tab:3}
\end{table*}


The stellar surface profile is then assembled using spherical bins of $R=\sqrt{X^{2}+Y^{2}}$, were we used 25, log-spaced concentric bins starting from the projected center of mass. From this profile, we fit a projected Plummer profile out to $R=4\times r_{1/2}$ in order to obtain the value of the effective radius, $R_{e}$. That is, the best-fit parameters are determined by adjusting the parameters of a projected Plummer in order to minimize a figure-of-merit function. The dispersion profiles are evaluate using the same bin spacing in spherical shells. The relative velocities found in projection are transformed to cylindrical values in correspondence to the coordinate system used in Section~\ref{sec:preliminaries}. That is, the velocity components parallel and transverse to radius $R$ follows $v_{\mathcal{R}} = (\boldsymbol{X}\cdot\boldsymbol{V})/R$ and $v_{\mathcal{T}} = |\boldsymbol{X}\wedge \boldsymbol{V}|/R$, respectively. In each bin shell of $R$, the relative velocities in projection are weighted by their associated stellar particle mass. Finally, the stellar mass-weighted velocity dispersions of the entire galaxy is measured within $4\times R_{e}$ for the value of $R_{e}$ determined from the surface density fit. We consider both random and streaming motions. 

\subsection{Results}
\begin{figure*}
    \centering
    \includegraphics[width=0.875\textwidth]{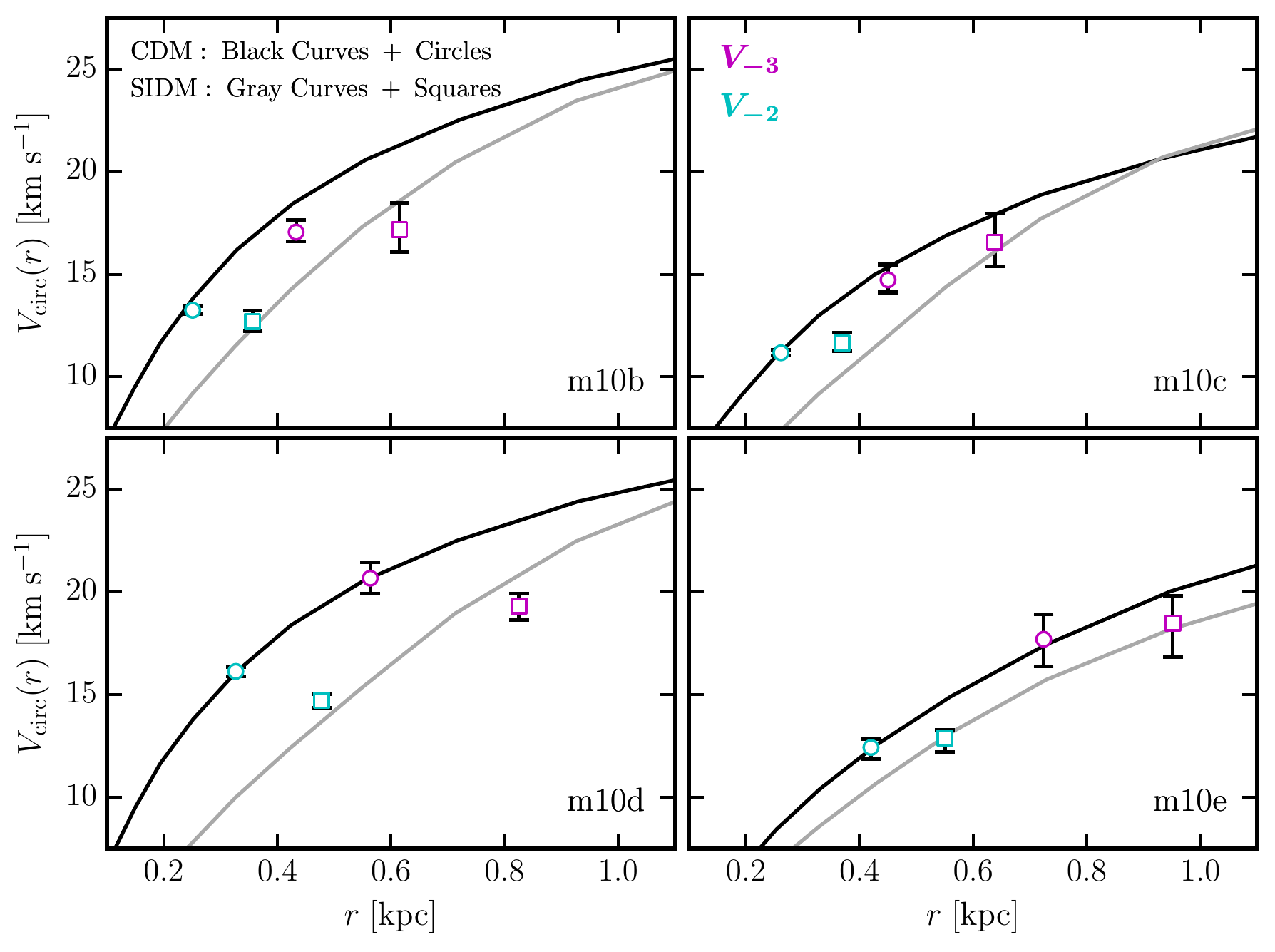}
    \caption{---
    {\bf \emph{Measurements from mock observations}}.
    The rotation curves of our galaxies compared to the estimators at the characteristic radii $r_{-2}$ and $r_{-3}$ of the stellar density profile. Black and gray lines show the rotation curves for each system simulated in CDM (cusps) and SIDM (cores). The estimators $V_{-2}$ (cyan) and $V_{-3}$ (magenta) are plotted at $r_{-2}$ and $r_{-3}$, respectively, where circles denote the estimators for the CDM galaxies and squares are for galaxies in SIDM. Error bars are the $1\sigma$ dispersion over all 1000 projections. Note that while the estimators are not perfect, they are accurate enough to discriminate between SIDM and CDM models in each case, especially when the two estimators are combined.  Estimates of the shapes of the rotation curves will be more uncertain than the overall circular velocity normalization at each radius.  Given large enough galaxy samples,  measurements should enable a strong discriminant between CDM and SIDM based on normalization alone.  
    }
    \label{fig:5}
\end{figure*}

Our key results are presented in both Table~\ref{tab:3} and Fig~\ref{fig:5}. In the table, we first list quantities measured to test the assumptions discussed in Section~\ref{sec:assumptions}. We start with columns 1 and 2, which give the log-gradient slope of the intrinsic radial velocity dispersion, $\gamma_{\sigma}$, at $r_{-2}$ and $r_{-3}$, respectively. These values are not precisely zero (as we have assumed in our idealized estimator) but they are small compared to the log-slope of the tracer profile (-3 and -2) at these radii and therefore are roughly in line with our assumptions. This behavior is found to be present for all of our galaxies, regardless of dark matter cores and cusps lying dormant. The radial anisotropy is similarly slowly varying though we have not summarized it here.

Columns 3 and 4 show ratios that measure the flatness of observable velocity profiles as the ratios $\widetilde{\sigma}_{\mathcal{T},-2} := \sigma_{\mathcal{T}}(r_{-2})/\langle \sigma_{\mathcal{T}}\rangle^{*}$ and $\widetilde{\sigma}_{\rm tot,-3} := \sigma_{\rm tot}(r_{-3})/(\sqrt{3}\langle \sigma_{\rm los}\rangle^{*})$, respectively. For the component transverse to the projected radius $R$, the median results are found to be well approximated by $\sigma_{\mathcal{T}}(r_{-2}) \simeq \langle \sigma_{\mathcal{T}}\rangle^{*}$ within $10\%$ even when considering the $68\%$ dispersion for galaxies either with cusps and cores. Interestingly, uncertainties are well constrained for all of the galaxies in our sample when just considering binned unit circles of projected radius $R$. Looking at the relation argued in \cite{wolf2010accurate} and here for the total intrinsic velocity dispersion (referring to column 4), the median results are found to be well approximated by $\sigma_{\rm tot}(r_{-3}) \simeq \sqrt{3}\langle \sigma_{\rm los}\rangle^{*}$ to better than about $10\%$ for the cusped galaxies. The galaxies with cores have this approximation accurate to $15-20\%$ when including the $1\sigma$ deviations.

Shown in Fig.~\ref{fig:5} are the actual circular velocity curves compared to the combined measurements of the estimators at their characteristic radii, $V_{-2}$ at $r_{-2}$ (cyan points) and $V_{-3}$ at $r_{-3}$ (magenta points). The vertical error bars of the estimators depict the $1\sigma$ dispersion from all 1000 mock projections. The total circular velocity profile is given by the black curves for the CDM and gray curves for SIDM. Columns 5 and 6 lists the ratio between these velocity estimators to the true value of the galaxies dynamical mass at the respective characteristic radii. The CDM galaxies perform remarkably well in predicting {\rm both} the actual circular velocity measurement at $r_{-2}$ and $r_{-3}$ within $10\%$ including uncertainties. The SIDM galaxies are as good to $20\%$ when including $1\sigma$ dispersions. By examining the outliers we see the worst offsets stem from difficulties in determining $r_{-2}$ and $r_{-3}$ of the simulated stellar density profile, as these profiles are, in essence, noisy, which makes the measurements of the log-gradient profiles without smoothing the density profile problematic. 

Since the idealized estimators, $V_{-2}$ and $V_{-3}$, predict the values of the dynamical profile to acceptable accuracy, we now see established predictions are when using characterizations modeled from the Plummer profile. Columns 9-12 in Table~\ref{tab:3} give the results for performing a fit using a Plummer profile on the projected surface density in each mock observation. The resulting values of $R_{e}$ are used to measure the stellar mass-weighted median dispersion, which have been depicted in Fig.~\ref{fig:5}. Columns 9 and 10 are the ratios of using the estimators with $R_{e}$ while columns 11 and 12 are the comparisons of the characteristic radii to the predicted mapping. We see that for many galaxies, the Plummer fits do not provide precise enough characterizations to infer the values of $r_{-2}$ and $r_{-3}$ to better than $\sim 20 \%$.  

As for modeling the slope of the underlying profile, we expect that the local inner-density behaves like a power-law, $\rho \propto r^{-\alpha}$ such that the integrated mass scales as $M \propto r^{3-\alpha}$. This leads us the expected behavior of circular velocity in relation to the local dark matter density: $V_{\rm circ}^{2} \propto r^{2-\alpha}$. We derive the implied slope given by the estimators by relating the inner density of the circular velocities as a power law that is defined like $\xi :=\Delta \log V_{\rm circ}/\Delta\log r$. This allows then to the relate the power laws for the density profile, i.e., $\alpha = 2(1-\xi)$.\footnote{\cite{di2013dependence} and \cite{tollet2016nihao} define cusps as $\alpha \simeq [1 - 1.5]$, which maps to $\xi \simeq 0.5$, and define cores as $\alpha \simeq [0 - 0.5]$, which maps to $\xi \simeq 0.75$.} The implied slope of the circular velocity estimators is given by the dashed red line in Fig.~\ref{fig:5}. In columns 7 and 8, we give the implied slope of the combined estimators, $\xi_{\rm est}$, and the true slope of the {\it dynamical} profile found at $r_{-2}$ and $r_{-3}$, $\xi_{\rm true}$, respectively. Without considering the $1\sigma$ dispersion of measurements, estimates from galaxies in CDM are predicted within $20 \%$ while the SIDM analogs are off by almost $50\%$. While the cuspy profiles are reasonably well measured, the SIDM core profiles estimated to be too cuspy via this method.  This is unfortunate, as  this precision is not enough to distinguish between a cusp and core.  However, the accuracy on the normalization ($V_{-2}$ at $r_{-2}$) is good enough to discriminate between absolute core densities expected for CDM vs. SIDM.  With large enough data sets, this will provide important constraints on models of this kind.  

\section{Discussion}
\label{sec:discussion}
We have used the spherical Jeans equation to infer two idealized mass estimators that depend on the stellar proper motions measured in observations. Specifically, we there are two radii, independent from one another, that potentially minimizes the anisotropy of the mass profile: one radius based off of measurements of the velocity dispersions {\it along} the line-of-sight and another radius from measurements for dispersions transverse along the plane of the sky. 


\subsection{Constraints from the Virial Theorem}
\label{sec:virial}
The scalar virial theorem has been historically utilized to provide approximate mass constraints for spheroidal galaxies \citep[e.g.][]{tully1997method,busarello1997virial}. 
 That is, the scalar virial theorem is  observationally applicable, such that dispersion-supported systems can probe the integrated mass profile within the stellar extent without the degeneracies provided by the anisotropy. 
It is constructed from the diagonalized components of the velocity dispersion tensor, which describes the {\it local} distribution of velocities at each point in space. The trace of diagonal components provide an extended scalar virial theorem \citep{errani2018virial}:
\begin{align}
    \langle \sigma_{\alpha}^{2} \rangle^{*}
    + \langle \sigma_{\delta}^{2} \rangle^{*}
    + \langle \sigma_{\rm los}^{2} \rangle^{*}
    &=
    4\pi G \int_{0}^{\infty}
    dr\ r n_{\star}(r)M(r)
    \label{eq:extended.SVT}
    \\
    &\equiv 
    \langle \sigma_{\rm tot}^{2} \rangle^{*}
    \nonumber 
    \, 
\end{align}
where $\langle \sigma_{\alpha}^{2} \rangle^{*}$ and $\langle \sigma_{\delta}^{2} \rangle^{*}$ are defined as the luminosity-averaged velocity dispersions of the two velocity components tangential to the line-of-sight. By design, Eq.~\eqref{eq:extended.SVT} provides a good integral constraint on the dynamical mass, as the entire expression is independent of the anisotropy. The line-of-sight component can be utilized as a constraint via the projected virial theorem \citep[e.g.][]{agnello2012core,errani2018virial}.  Adding dispersions in the $\alpha$ and $\delta$ directions  would enable a tighter constraint on $\langle \sigma_{\rm tot}^{2} \rangle^{*}$.  Note however, that when written this way we do not provide any additional constraint on $\beta$. 

Working in a Cartesian coordinate system, such that  $\mathrm{los} \rightarrow z $, $\alpha \rightarrow x$ and $\delta \rightarrow y$, then spherical symmetry would demand each component of velocity dispersion be equal: $\langle \sigma_{x}^{2} \rangle^{*} = \langle \sigma_{y}^{2} \rangle^{*} = \langle \sigma_{z}^{2}\rangle^{*}$.  The coordinate system introduced in Section~\ref{sec:preliminaries} does not force this equality and allows for separate components of the luminosity-averaged velocity dispersion to constrain the velocity dispersion anisotropy $\beta$ \citep{strigari2007astrometry}. The two components, $\sigma_{\mathcal{R}}$ and $\sigma_{\mathcal{T}}$, depend on $\beta$ differently and are not necessarily equal.\footnote{Though symmetry demands $\langle \sigma_{\mathcal{T}}^{2} \rangle^{*} + \langle \sigma_{\mathcal{R}}^{2} \rangle^{*} = \frac{2}{3} \langle \sigma_{\rm tot}^{2} \rangle^{*} = 2 \langle \sigma_{\rm los}^{2} \rangle^{*}$.} Note that when Eqs.~\eqref{eq:sig_los.mapping},~\eqref{eq:sig_projR.mapping},~and~\eqref{eq:sig_projT.mapping} are added together we find $\langle \sigma_{\mathcal{T}}^{2} \rangle^{*}
+ \langle \sigma_{\mathcal{R}}^{2} \rangle^{*}
+ \langle \sigma_{\rm los}^{2} \rangle^{*} = \langle \sigma_{\rm tot}^{2} \rangle^{*}$ such that Eq.~\eqref{eq:extended.SVT} can be satisfied.  By examining the components separately we can have mass estimators that provide information at a different radius than the one enabled from line-of-sight motions alone.

\subsection{Possible Biases in Jeans Modeled Mass Estimates}
Our mass estimates rely on the fact that dispersion-supported systems are approximately in dynamical equilibrium and are accurately modeled by the spherical Jeans equation. Non-steady-state systems, ones that significantly deviate from dynamical equilibrium, can lead to biased estimations of the complete dynamical mass. This can lead to systematically biased mass estimates \citep[e.g.][]{amorisco2011phase,errani2018virial}. For the simulated galaxy sizes considered in our analysis, mass estimates with short time-scale fluctuations of the potential well are non-trivially biased \citep{elbadry2017jeans,gonzalez2017dwarf}. For the largest kind of dispersion-supported systems, ones with a stellar mass of $M_{\star} \approx 10^{8-10} M_{\odot}$, uncertainty are as large $20 \%$ of the dynamical mass. To minimize the variability of energetic outflows, mass estimates are best focused on dwarf spheroidals at around the threshold of lowest detectability, i.e., low-mass dwarfs, as this should reduce the likelihood of potential fluctuations biasing the stellar tracers. Using simulated data, we have shown that our estimate at $r_{-2}$ is able to obtain the normalization to better than $20\%$ when using $V_{-2}$ for low-mass dwarf galaxies. As for the applicability to observations, it is import that careful measurements of the highest precision are obtained in order to dissociate between possible models embellished with systematic errors 

Although our simulated galaxies are analogous to those in the field, local group satellites also experience tidal stripping of the main halo, which can preferentially bias the estimates of the satellites dynamical mass. However, analysis from \cite{klimentowski2007tidal} has already eluded that velocity dispersions are well modeled by the Jeans equation for even in the case of mildly tidally disrupted dwarf galaxies, as long as unbound, interloping stars are properly accommodated for in the stellar sample. For the case of Draco and Sculptor considered here, they are both satellites of the Milky Way and are therefore, in principle, subjected to tidal forces that could render mass models from the Jeans equation inadequate. However, no sign of strong tidal influence is apparent \citep{piatek2002draco,coleman2005tidal}. 

\section{Concluding Remarks}
\label{sec:conclusion}
Using the spherical Jeans equation, we have derived a mass estimator that depends on stellar kinematics measured along the plane of the sky, specifically the velocity dispersion tangential to the projected radius $R$. We have shown that under idealized but reasonable assumptions, Eq.~\eqref{eq:lazar_mass} provides the cumulative mass within a characteristic radius, $r_{-2}$,  independent of the stellar velocity dispersion anisotropy $\beta$.  This ideal radius is where the log-slope of the underlying tracer profile is $-2$.  For Plummer profiles $r_{-2} \simeq 4R_{e}/5 \simeq 3r_{1/2}/5$.  We also showed that a $\beta$-independent estimator does not exist for the velocity dispersion parallel along the plane of the sky. Fig.~\ref{fig:1} summarizes this result.
Our derivation followed the approach in \cite{wolf2010accurate}, and relied on similar assumptions: that the stellar velocity dispersion profiles $\sigma_{r}(r)$ and $\beta(r)$ vary slowly compared to the tracer profile itself out to $r_{1/2}$. 

To test our assumptions and our estimators, we employ previously-published simulations of dwarf galaxies done for both CDM and SIDM dark matter physics. We find that $\sigma_{\mathcal{T}}$ is indeed flat in the vicinity of $r_{-2}$ for both dwarf galaxies of CDM and SIDM and found that our mass estimator is accurate in quantifying the enclosed mass at $r_{-2}$. For CDM, the estimates for the dynamic rotation curves are found to be accurate to $10\%$ for both estimators while SIDM are accurate to $15\%$. This level of absolute mass accuracy is good enough to discriminate between expected core densities in SIDM and CDM models.  Unfortunately, this level of accuracy is {\em not} good enough to regularly measure slopes at the precision required to differentiate between cusps and cores in real data without deeper prior to help us understand the underlying tracer profile shape in real galaxies. However, the difference in absolute circular velocity predicted between SIDM and CDM at these radii is well within the normalization uncertainties of the estimators (see Fig \ref{fig:5}).

As an example of the applicability of our estimator, we have combined it with the \cite{wolf2010accurate} estimator at $r_{-3}$ for line-of-sight velocities to explore the mass profiles of Draco and Sculptor. Both galaxies are consistent with inhabiting cuspy NFW subhalos with densities consistent with those expected in CDM with $V_{\rm max} \simeq 34$ and $19\ \rm km\ s^{-1}$, respectively, though current uncertainties allow for a variety of inner profile slopes and are consistent with SIDM densities given the sparsity of the data.  In the coming era of precision-based measurements of stellar proper motions, we expect the internal structure of dwarf galaxies to be revealed with more clarity.

\section*{Acknowledgements}
We are thankful to the anonymous referee for their invaluable feedback that helped improve the earlier version of this article. We would like to thank Victor Robles and Alex Fitts for facilitating access to their simulations.
We are thankful to Michael Boylan-Kolchin and Josh Simon for comments on early versions of the article. 
Lazar and Bullock are supported by the NSF grant AST-1910965 .
The analysis in this article made extensive use of the python packages \texttt{NumPy} \citep{van2011numpy}, \texttt{SciPy} \citep{oliphant2007python}, and \texttt{Matplotlib} \citep{hunter2007matplotlib}; We are thankful to the developers of these tools. This research has made all intensive use of NASA's Astrophysics Data System (\url{https://ui.adsabs.harvard.edu/}) and the arXiv eprint service (\url{http://arxiv.org}).

\bibliography{references}

\appendix

\section{Mass Profiles as a Function of Observables}
\label{sec:mass.profiles}
Here we derive a single expression for the mass profile of spheroidal galaxies as a function of observable combinations found in the plane of the sky. A crucial argument that we have imposed previously is that projected observables can be re-formalized to de-projected quantities, and vice-versa. To do this we make note of the utilization of the Abel inversion \citep{binney2011galactic}:
\begin{align}
    f(x) = \int_{x}^{\infty} \frac{dt}{\sqrt{t-x}}g(t)
    \iff
    g(x) = -\frac{1}{\pi} \int_{t}^{\infty} \frac{dx}{\sqrt{x-t}} \frac{df}{dt}
    \, .
    \label{eq:abel.inversion}
\end{align}

\begin{figure*}
    \centering
    \includegraphics[width=0.925\columnwidth]{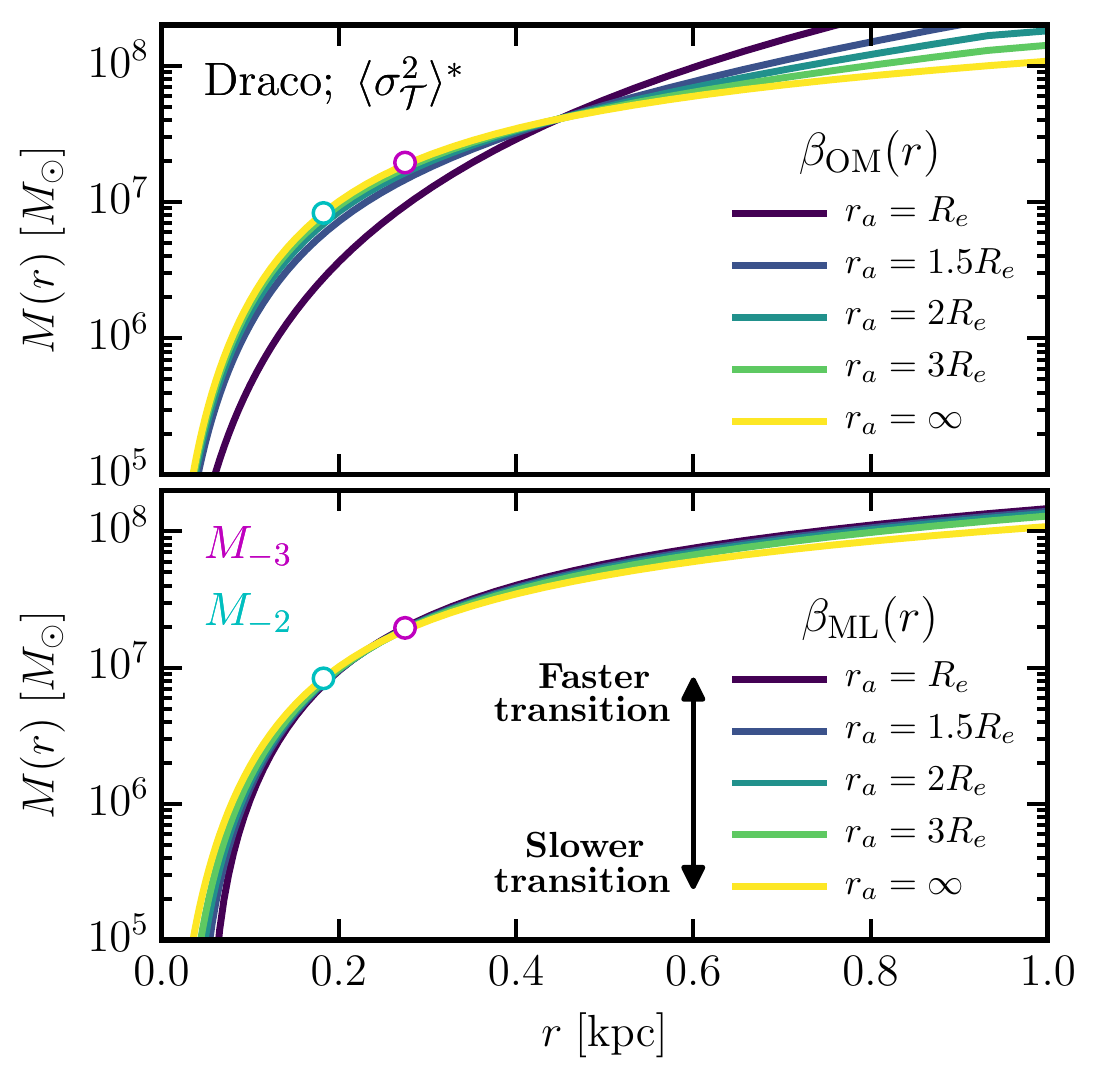}
    \includegraphics[width=0.925\columnwidth]{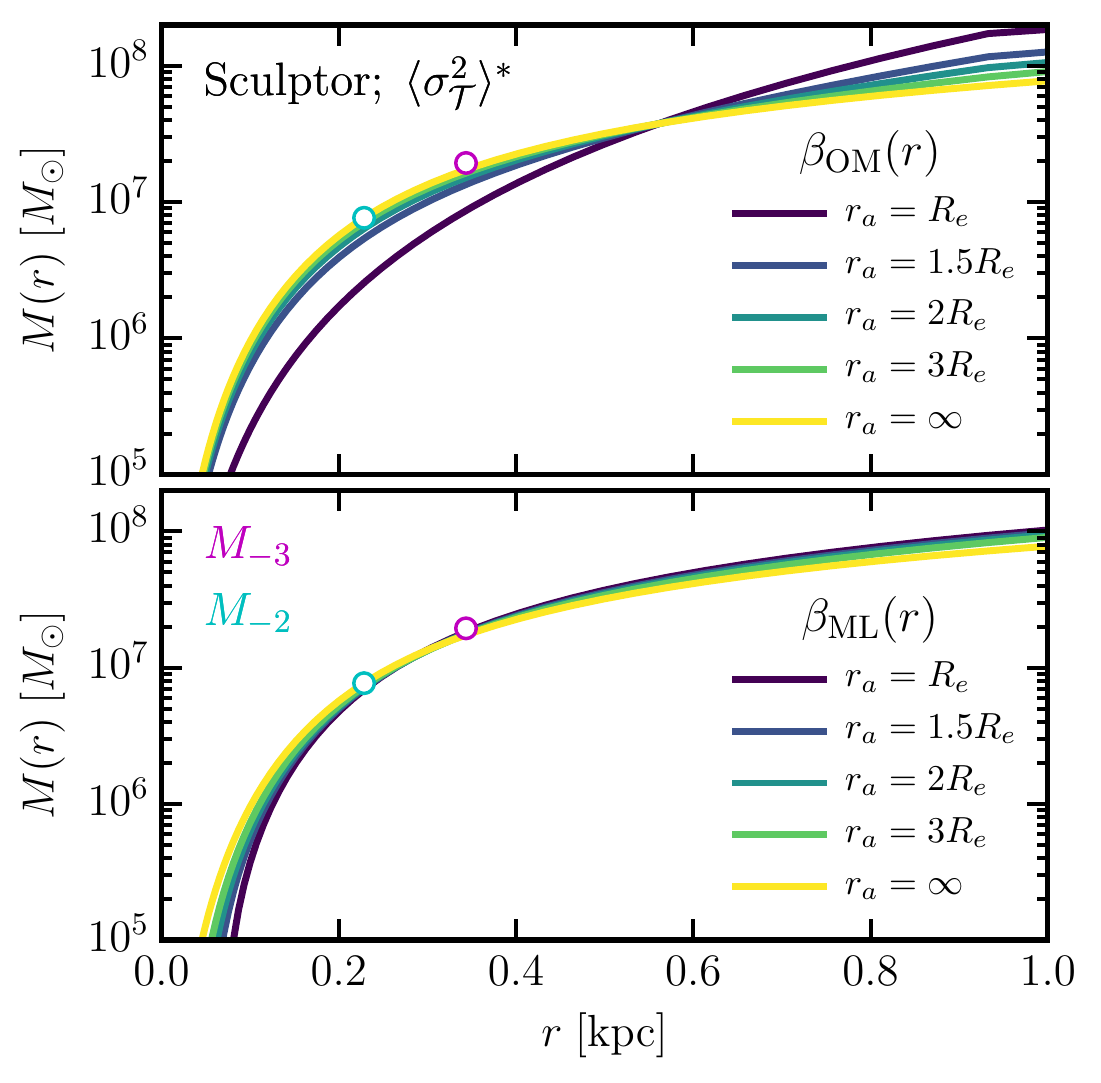}
    \caption{---
        {\bf \emph{Integrated mass profiles for Draco and Sculptor with radial anisotropy}}.
        The lines correspond to several choices of fixed velocity dispersion anisotropy $\beta_{0}$. Mass profile curves are generated using Eq.~\eqref{eq:projT.mass.profile.general} assuming a Plummer profile and constant $\sigma_{\rm los}$. Here, the small white circle indicates the mass predicted by the {\it idealized} estimator, Eq.~\eqref{eq:lazar_mass}, which intersects at the mass that is independent of the value of $\beta_{0}$. 
    }
    \label{fig:A1}
\end{figure*}

\subsection{Measurements along the Line-of-sight}
While a complete derivation is given in \cite{wolf2010accurate}, we quote the mass profile for completeness: Given measurements along the line-of-sight and assuming a constant anisotropy model, $\beta(r) = \beta_{0}$, the mass profile come to be
\begin{align}
    M(r|\beta_{0})
    &=
    \frac{\Big\{ \mathcal{K}_{1}(r,R|\beta_{0}) + \mathcal{K}_{2}(r,R|\beta_{0}) \Big\}}{G\pi (\beta_{0} - 1) n_{\star}(r)}
    \label{eq:los.mass.profile.constant}
    \, ,
\end{align}
where the integral kernels are
\begin{align*}
    \mathcal{K}_{1}(r,R|\beta_{0})
    &=
    \int_{r^{2}}^{\infty} 
    dR^{2}\ R^{2} \frac{d^{2} (\Sigma_{\star}\sigma_{\rm los}^{2})}{(dR^{2})^{2}} 
    \frac{2r^{3}/R^{3}}{\sqrt{1-r^{2}/R^{2}}} 
    \\
    \mathcal{K}_{2}(r,R|\beta_{0})
    &=
    \int_{r^{2}}^{\infty} 
    dR^{2}\ R^{2} \frac{d^{2} (\Sigma_{\star}\sigma_{\rm los}^{2})}{(dR^{2})^{2}} 
     \beta_{0}\frac{3-2\beta_{0}}{\beta_{0}-1} 
    \\
    &\qquad \times 
    \left(\frac{r}{R}\right)^{\frac{1}{1-\beta_{0}}}
    \mathcal{B}_{1-r^{2}/R^{2}} \left( \frac{1}{2} , \frac{2-3\beta_{0}}{2(1-\beta_{0})} \right) 
    \nonumber 
    \, ,
\end{align*}
For compactness, the lower-incomplete beta function is incorporated:
\begin{align}
    \mathcal{B}_{x}(p,q) := \int_{0}^{x} dy\ y^{p-1}(1-y)^{q-1}  
    \, .
\end{align}
In Fig.~\ref{fig:1}, we demonstrate the robustness of this mass profile. As an example, we consider the assumptions used in deriving the idealized case of $M_{-3}$: a constant $\beta$ model and a constant velocity dispersion measurement $\sigma_{\rm los}^{2}(R) \simeq \langle\sigma_{\rm los}^{2}\rangle^{*}$.  For $\langle\sigma_{\rm los}^{2}\rangle^{*}$, we use the median observational quantities for Draco and Sculptor given in Table~\ref{tab:1}. As shown, the different values of $\beta_{0}$ converge at $r_{-3}$ of the stellar density profile, at least for an assumed Plummer profile. The idealized mass estimator, $M_{-3}$, is shown as the white dot intersecting where the profiles converge.

\subsection{\emph{Plane of the sky}: Measurements Parallel to \emph{R}}
\label{sec:mass.profile.projR}
To further clarify that lack of a radius that minimizes the anisotropy in an idealized case, consider the utilization of the Jeans equation for measurements of $\sigma_{\mathcal{R}}$. We start by the massaging the form of Eq.~\eqref{eq:sig_projR.mapping} in order to isolate out the $R$ dependence in the integral kernel:
\begin{align}
    \Sigma_{\star} \sigma_{\mathcal{R}}^{2}(R)
    &=
    \int_{R^{2}}^{\infty}
    \frac{dr^{2}}{\sqrt{r^{2} - R^{2}}}
    \left[
    1 - \beta(r) + \frac{R^{2}}{r^{2}}\beta(r)
    \right]
    n_{\star}\sigma_{r}^{2}(r)
    \\ \nonumber 
    &= 
    \int_{R^{2}}^{\infty}\frac{dr^{2}}{\sqrt{r^{2} - R^{2}}}n_{\star}\sigma_{r}^{2}
    -
    \int_{R^{2}}^{\infty}dr^{2}\frac{n_{\star}\sigma_{r}^{2}
    \left(
     r^{2} - R^{2}
    \right)}{\sqrt{r^{2} - R^{2}}}
    \\ \nonumber
    &=
    \int_{R^{2}}^{\infty}\frac{dr^{2}}{\sqrt{r^{2} - R^{2}}}    
    \Big[ 
    n_{\star}\sigma_{r}^{2}(r) - \int_{r^{2}}^{\infty} d\tilde{r}^{2} \frac{\beta n_{\star}\sigma_{r}^{2}}{2 \tilde{r}^{2}}
    \Big]
    \, .
\end{align}
Here in the second line we expanded the second term with integration by parts and evaluating the boundary integration to null by motivating that combination of $\beta n_{\star} \sigma_{r}^{2}$ falls faster than $r^{-1}$ at large $r$.

We then use the invertable form and deproject via an Abel inversion to obtain
\begin{align}
    n_{\star}\sigma_{r}^{2}(r) - \int_{\log r}^{\infty} d\log \tilde{r}\ \beta n_{\star} \sigma_{r}^{2}
    &=
    -\frac{1}{\pi} \int_{r^{2}}^{\infty} \frac{dR^{2}}{\sqrt{R^{2} -r^{2}}} 
    \frac{d(\Sigma_{\star}\sigma_{\mathcal{R}}^{2})}{dR^{2}}
     \, .
\end{align}
To isolate out $n_{\star} \sigma_{r}^{2}$, we differentiate with respect to $\log r$,
\begin{align}
    \frac{d(n_{\star}\sigma_{r}^{2})}{d\log r} 
    + \beta n_{\star} \sigma_{r}^{2}
    &=
    -\frac{2r^{2}}{\pi} \int_{r^{2}}^{\infty} \frac{dR^{2}}{\sqrt{R^{2} -r^{2}}} 
    \frac{d^{2}(\Sigma_{\star}\sigma_{\mathcal{R}}^{2})}{(dR^{2})^{2}} 
    \, ,
\end{align}
and then deploy the integrating factor
\begin{align}
    h(r) = \exp\left\{ \int_{\log a}^{\log r} d\log\tilde{r}\ \beta(r)  \right\}
    \, ,
\end{align}
where $a$ is a constant chosen so that the value of the integrand approaches zero at the lower limit. This then gives us
\begin{align}
    n_{\star}\sigma_{r}^{2}(r|\beta)
    &=
    -\frac{h^{-1}}{\pi} \int_{r^{2}}^{\infty} d\tilde{r}^{2}
    \bigg[ 
    \int_{\tilde{r}^{2}}^{\infty} \frac{dR^{2}}{\sqrt{R^{2}-r^{2}}}
    \frac{d^{2}(\Sigma_{\star}\sigma_{\mathcal{R}}^{2})}{(dR^{2})^{2}} 
    \bigg]h
    \nonumber \\
     &=
    -\frac{h^{-1}}{\pi} 
    \int_{r^{2}}^{\infty} dR^{2}\ 
    \bigg[ 
    \int_{r^{2}}^{R^{2}} \frac{d\tilde{r}^{2}}{\sqrt{R^{2}-\tilde{r}^{2}}} h
    \bigg]
    \frac{d^{2}(\Sigma_{\star}\sigma_{\mathcal{R}}^{2})}{(dR^{2})^{2}} 
    \, .
\end{align}
Here, $n_{\star}\sigma_{r}^{2}(r)$ can be modeled by the adoption of a parametric form of $\beta(r)$. This can also be taken and inserted in Eq.~\eqref{eq:jeans_mass} to model the integrated mass.

For our idealized case, suppose the anisotropy is taken to be a constant value for the inner region of the system, $\beta(r) = \beta_{0}$. We will have $h(r) \rightarrow r^{\beta_{0}}$ which allows us rewrite the inner integral in terms of the lower-incomplete beta function:
\begin{align}
    n_{\star}\sigma_{r}^{2}(r|\beta_{0})
    &=
    \frac{r^{-\beta_{0}}}{\pi} 
    \int_{r^{2}}^{\infty} dR^{2}\ R^{\beta_{0} + 1}
    \frac{d^{2}(\Sigma_{\star}\sigma_{\mathcal{R}}^{2})}{(dR^{2})^{2}} 
    \mathcal{B}_{1-r^{2}/R^{2}}\left( \frac{1}{2}, \frac{\beta_{0} + 2}{2} \right)
    \, .
\end{align}

The mass profile is then obtained by hitting the previous expression with a derivative in respect to $\log r$ and insert it into Eq.~\eqref{eq:jeans_mass} to acquire the implied profile
\begin{align}
    M(r|\beta_{0}) &=
    \frac{r}{\pi G n_{\star}(r)}
    \Big\{ \widetilde{\mathcal{R}}_{1}(r|\beta_{0}) +  \widetilde{\mathcal{R}}_{2}(r|\beta_{0}) \Big\}
    \label{eq:projR.mass.profile.constant}
    \, ,
\end{align}
where
\begin{align*}
    \tilde{\mathcal{R}}_{1}(r|\beta_{0}) &=
     2 r^{2}\int_{r^{2}}^{\infty} \frac{dR^{2}}{\sqrt{R^{2} - r^{2}}}
    \frac{d^{2}(\Sigma_{\star}\sigma_{\mathcal{R}}^{2})}{(dR^{2})^{2}} 
    \, ,
    \\
    \tilde{\mathcal{R}}_{2}(r|\beta_{0}) &=- \beta_{0} r^{-\beta_{0}} \int_{r^{2}}^{\infty} dR^{2}
    R^{\beta_{0} + 1} \frac{d^{2}(\Sigma_{\star}\sigma_{\mathcal{R}}^{2})}{(dR^{2})^{2}} 
    \mathcal{B}_{1-r^{2}/R^{2}}\left( \frac{1}{2}, \frac{\beta_{0} + 2}{2} \right) 
    \, .
\end{align*}
This relation replaces the dependence of deriving the mass of a dispersion-supported system from unknown radial velocity dispersion with the second ordered derivatives of the the observable product, $\Sigma_{\star}\sigma_{\mathcal{R}}^{2}(R)$.  The middle plot in Fig.~\ref{fig:1} realizes Eq.~\eqref{eq:projR.mass.profile.constant} for various values of $\beta_{0}$ and a constant velocity dispersion, $\sigma^{2}_{\mathcal{R}} \simeq \langle \sigma_{\mathcal{R}}^{2}\rangle^{*}$. We use the medium parameters of Sculptor given in Table~\ref{tab:1} as demonstration. As we predicted previously, none of the constant $\beta_{0}$ dependent Jeans mass profiles converge to a mass value like we have seen in in the top plot in Fig.~\ref{fig:1}. 

\subsection{\emph{Plane of the sky}: Measurements Transverse to \emph{R}}
Consider the application of the Jeans equation for measurements based on the mapping of Eq.~\eqref{eq:sig_projT.mapping}. Since this is already in a form that is invertable, we deproject via an Abel inversion in order to isolate out $n_{\star}\sigma_r^{2}(r)$ combination:
\begin{align}
    n_{\star}\sigma_{r}^{2}(r)
    &=
    \frac{1}{\pi(\beta -1)}\int_{r^{2}}^{\infty}
    \frac{dR^{2}}{\sqrt{R^{2} - r^{2}}}
    \frac{d(\Sigma_{\star} \sigma_{\mathcal{T}}^{2})}{dR^{2}}
    \, .
    \label{eq:sigr.profile.projT}
\end{align} 
We see that the above relation is unique in comparison with what we have seen in the previous sections, as Eq.~\eqref{eq:sigr.profile.projT} is unembelished and relatively simple in its form. This allows us to write a mass profile that can be dependent on some arbitrary form of $\beta(r)$.

\subsubsection{Spatially Constant Velocity Dispersion Anisotropy}
Assume a constant anisotropy model, $\beta(r) = \beta_{0}$. It is straightforward to differentiate both sides with respect to $\log r$ and massage it to the form that, along with Eq.~\eqref{eq:sigr.profile.projT}, can be inserted into Eq.~\eqref{eq:jeans_mass} to obtain the implied mass profile
\begin{align}
    M(r|\beta_{0})
    &=
    \frac{2r}{G\pi(1-\beta_{0})n_{\star}(r)}
    \Big\{ \widetilde{\mathcal{T}}_{1}(r|\beta_{0}) + \widetilde{\mathcal{T}}_{2}(r|\beta_{0})\Big\}
    \, , 
    \label{eq:projT.mass.profile.constant}
\end{align}
where
\begin{align*}
    \widetilde{\mathcal{T}}_{1}(r|\beta_{0})
    &=
    r^{2}\int_{r^{2}}^{\infty} 
    \frac{dR^{2}}{\sqrt{R^{2} - r^{2}}}
    \frac{d^{2}(\Sigma_{\star} \sigma_{\mathcal{T}}^{2})}{(dR^{2})^{2}}
    \, ,
    \\
    \widetilde{\mathcal{T}}_{2}(r|\beta_{0})
    &=
    \beta_{0}
    \int_{r^{2}}^{\infty} 
    \frac{dR^{2}}{\sqrt{R^{2} - r^{2}}}
    \frac{d(\Sigma_{\star} \sigma_{\mathcal{T}}^{2})}{dR^{2}}
    \, .
\end{align*}
With this, we have eliminated the dependency of the unknown dispersion profile, $\sigma_{r}(r)$, and have the mass profile be dependent only on the form of well-defined observables and an arbitrary value of $\beta_{0}$. The bottom plot in Fig.~\ref{fig:1}, actualizes this mass profile by considering the idealized assumptions used in the derivation of $M_{-2}$: an constant anisotropy model and a constant velocity dispersion $\sigma_{\mathcal{T}}^{2} \simeq \langle \sigma_{\mathcal{T}}^{2}\rangle^{*}$. We use the median values of Scupltor from Table~\ref{tab:1} for as a demonstration. Using our idealized conditions, we recover the {\it idealized} results derived previously, where the of the mass profiles converge at $r_{-2}$ for various values of constant $\beta_{0}$.

\subsubsection{General Velocity Dispersion Anisotropy}
Assume now a generalized $\beta(r)$ model, we differentiate both sides of  Eq.~\eqref{eq:sigr.profile.projT} with respect to $\log r$ to have
\begin{align}
    \frac{d (n_{\star}\sigma_{r}^{2})}{d\log r}(\beta-1) - \gamma_{\beta} \beta n_{\star} \sigma_{r}^{2}(r)
    &=
    \frac{2r^{2}}{\pi} 
    \int_{r^{2}}^{\infty} \frac{dR^{2}}{\sqrt{R^{2} - r^{2}}} 
    \frac{d^{2}(\Sigma_{\star} \sigma_{\mathcal{T}}^{2})}{(dR^{2})^{2}}
    \, .
\end{align}
We then expand out the differential on the left-hand side and substitute its resulting form into Eq.~\eqref{eq:jeans_mass} to obtain the implied mass profile for some generalized $\beta(r)$ model:
\begin{align}
    M(r|\beta)
    &=
    \frac{2r}{G\pi[1-\beta(r)]n_{\star}(r)}
    \left\{ \widetilde{\mathcal{T}}_{1}(r|\beta) + \widetilde{\mathcal{T}}_{2}(r|\beta) \right\}
    \, , 
    \label{eq:projT.mass.profile.general}
\end{align}
where
\begin{align*}
    \widetilde{\mathcal{T}}_{1}(r|\beta)
    &=
    r^{2} \int_{r^{2}}^{\infty} \frac{dR^{2}}{\sqrt{R^{2} - r^{2}}} 
    \frac{d^{2}(\Sigma_{\star} \sigma_{\mathcal{T}}^{2})}{(dR^{2})^{2}}
    \, ,
    \\
    \widetilde{\mathcal{T}}_{2}(r|\beta)
    &=
    \beta(r) 
    \left[ 1 - \frac{\gamma_{\beta}}{2[1-\beta(r)]} \right]
    \int_{r^{2}}^{\infty} \frac{dR^{2}}{\sqrt{R^{2} - r^{2}}} 
    \frac{d(\Sigma_{\star} \sigma_{\mathcal{T}}^{2})}{dR^{2}}
    .
\end{align*}
In the limiting case of a constant $\beta_{0}$ model, we see that we re-obtain Eq.~\eqref{eq:projT.mass.profile.constant}. 
In particular, $\gamma_{\beta}$ term, if large enough, could nominally shift the location of $r_{-2}$ of the implied mass profile, as this is true for any anisotropic characterizations that become close to isotropic. However, the behavior of the $\beta \gamma_{\beta}$ combination in $\widetilde{\mathcal{T}}_{2}(r)$ is still well behaved.

In Fig.~\ref{fig:A1}, we demonstrate Eq.~\eqref{eq:projT.mass.profile.general} by only assuming a constant velocity dispersion, $\sigma_{\mathcal{T}}^{2}(R) \simeq \langle \sigma_{\mathcal{T}}^{2}\rangle^{*}$, and a Plummer profile. We consider two functional forms frequently used to model in observational Jeans modeling studies: 
\begin{itemize}
    \item 
    The {\it Osipkov-Merritt} \citep[][subscript OM]{osipkov1979ellipsoidal,merritt1987virial} profile is formulated from a one-parameter family of spherically stellar systems, such that,
    \begin{align}
        \beta_{\rm OM}(r)
        &= 
        \frac{r^{2}}{r^{2} + r_{a}^{2}}
        \, ,
    \end{align}
    where $r_{a}$ is the {\it anisotropy} radius that gives the stellar sub-component of a velocity distribution. Notably, $\beta \rightarrow 0$ as $r\ll r_{a}$ and $\beta \rightarrow 1$ as $r\gg r_{a}$. Different values of $r_{a}$ are given in the top row.
    \vspace{1.5ex}

    \item
     Another parametrization is introduced by {\it Mamon-\L okas} \citep[][subscript ML]{mamon2005virial} profile:
    \begin{align}
        \beta_{\rm ML}(r)
        &= 
        \frac{1}{2}
        \frac{r}{r + r_{a}}
        \, .
    \end{align}
    Similar limits as $\beta_{\rm OM}(r)$, but $\beta \rightarrow 1/2$ as $r \gg r_{a}$. Different values of $r_{a}$ are given in the bottom row.
\end{itemize}

The values of $r_{a}$ are subjugated between two extreme cases, one where $r_{a} = \infty$, making the profile isotropic at all radii, and another as low as $r_{a} = R_{e}$, which allows for a quicker transition of the profile being anisotropic to isotropic around the region of $R_{e}$. For values of $r_{a} \gtrapprox 2R_{e}$ the estimator appears robust for both models of $\beta_{\rm OM}(r)$ and $\beta_{\rm ML}(r)$. In the instances where $r_{a} \simeq R_{e}$, both $M_{-3}$ (magenta circle) and
$M_{-2}$ (cyan circle) fail for the faster transition value of $\beta_{\rm OM}$ while is still consistent with $\beta_{\rm ML}$. Indubitably, as $M_{-3}$ and $M_{-2}$ breaks down as the log-slope of the anisotropy profile is too large. The relative simplicity of Eq.~\eqref{eq:projT.mass.profile.general} allows for a complete modeling of the mass profile based off of the proper motions tangential along the plane of the sky. The caveat here is that realization of the complete dynamical profile requires highly accurate observational data, as first and second derivatives of surface profile is required. 

\bsp	
\label{lastpage}
\end{document}